\documentclass[twocolumn,twoside]{IEEEtran}
\ifCLASSINFOpdf
  
\else
 
\fi

\ifCLASSOPTIONcompsoc
    \usepackage[caption=false, font=normalsize, labelfont=sf, textfont=sf]{subfig}
\else
    \usepackage[caption=false, font=normalsize]{subfig}
\fi
\usepackage{lipsum}%

\usepackage{balance}
\usepackage{multicol}   
\usepackage{cite}
\usepackage{gensymb}
\usepackage{multirow}
\usepackage{graphics}  
\usepackage{epsfig} 
\usepackage{graphicx}
\usepackage{epstopdf}
\usepackage{textcomp}
\usepackage{amsmath}
\usepackage{mathtools}
\usepackage{xfrac}
\usepackage{lipsum,color}
\usepackage{mathtools,tikz,caption}
\usepackage{amssymb}
\usepackage{amsfonts}
\usepackage{float,stfloats,blindtext}
\usepackage{resizegather}
\usepackage{hyperref}
\begin{document}
	
\title{\textcolor{black}{CubeSat-Enabled Free-Space Optics: \\Joint Data Communication and Fine Beam Tracking}}

\author{Hossein Safi, Mohammad Taghi Dabiri, Julian Cheng, \textit{Fellow}, \textit{IEEE}, \\Iman Tavakkolnia, \textit{Member}, \textit{IEEE}, and    Harald Haas, \textit{Fellow}, \textit{IEEE} 
   
	\thanks{ H. Safi, I. Tavakkolnia, and H. Haas are with the Electrical Division, Department of Engineering, University of Cambridge, Cambridge, UK, e-mails: \{hs905, it360, huh21\}@cam.ac.uk. M. T. Dabiri is with the Department of Electrical Engineering, Qatar University, Doha, Qatar, e-mail: (m.dabiri@qu.edu.qa). J. Cheng is with the School of Engineering,
		The University of British Columbia, Kelowna, BC,
		Canada (e-mail: julian.cheng@ubc.ca). \\This work is a contribution by Project REASON, a UK Government funded project under the Future Open Networks Research Challenge (FONRC) sponsored by the Department of Science Innovation and Technology (DSIT).}}

\maketitle

\begin{abstract}
The integration of CubeSats with Free Space Optical (FSO) links accelerates a major advancement in high-throughput, low-Earth orbit communication systems. However, CubeSats face challenges such as size, weight, and power (SWaP) limitations, as well as vibrations that cause fluctuations in the angle-of-arrival (AoA) of the optical beam at the receiver. These practical challenges make establishing CubeSat-assisted FSO links complicated. To mitigate AoA fluctuations, we expand the receiver’s field of view and track the location of the focused beam spot using an array of avalanche photodiodes at the receiver. Initially, we model the optical channel between the transmitter and the detector array. Furthermore, to reduce the computational load of maximum likelihood sequence detection, which is infeasible for CubeSats due to SWaP constraints, we propose a sub-optimal blind sequence data detection approach that relies on the generalized likelihood ratio test (GLRT) criterion. We also utilize combining methods such as equal gain combining (EGC) and maximal ratio combining (MRC) for data detection, benchmarking their performance against the GLRT-based method. Numerical results demonstrate that the proposed low-complexity GLRT-based method outperforms the combining methods, achieving performance close to that of the ideal receiver.
	

\end{abstract}


\IEEEpeerreviewmaketitle
\vspace{-2.5 mm}
\section{Introduction}
\label{I}

\IEEEPARstart{I}{n} the sixth-generation (6G) era, the race to offer and commercialize high data rate services faces challenges due to limited radio frequency (RF) spectrum availability. As a result, there is a growing interest in  exploiting large blocks of spectrum in higher frequency bands to accommodate the need for increased peak rates and capacity in wireless communication links \cite{ITU-R_M2160}. Free-space optical (FSO) communication has emerged as a promising solution to address these challenges and enhance modern communication systems \cite{FSO6G}. Unlike RF communication, FSO uses optical wavelengths to transmit data through free space, offering advantages such as low power consumption, significantly higher unlicensed bandwidth, and immunity to electromagnetic interference. In satellite communication, where traditional commercial bandwidth allocations (i.e., S, X, Ku, and Ka bands) struggle to meet the demands of data-intensive 6G and beyond applications \cite{9210567}, FSO technology offers the potential for higher data rates, enhanced security, and reduced size, weight and power (SWaP) due to smaller antenna requirements \cite{7553489}. These characteristics make FSO-equipped satellite systems well-suited for a wide range of applications \cite{9219130}.

Interestingly, the space industry has recently witnessed a remarkable shift from traditional large and costly spacecraft to smaller, more accessible missions \cite{hassan2020dense}. This transformation was catalyzed by advancements in commercial-off-the-shelf (COTS) technology, which enabled the miniaturization of spacecraft components \cite{saeed2020cubesat}. Accordingly, a class of nanosatellites, also known as CubeSats, has emerged as a cost-effective platform. Their compact size, low cost, and rapid development timelines have led to their adoption for a variety of applications. CubeSats typically operate in low earth orbit (LEO) to offer low latency communication, reduced lunch cost, and to support rapid response missions. They also adhere to standardized dimensions, measured in ``units'' (U), where 1U corresponds to a cube measuring 10 $\times$ 10 $\times$ 10 centimetres. Notably, the global CubeSat market, valued at USD 461 million in 2023, is projected to reach USD 1445.04 million by 2031, with a compound annual growth rate of 15.34\% \cite{Skyquest}. Consequently, major companies like Google, SpaceX, OneWeb, and Facebook have shown interest in CubeSats for various applications, including earth monitoring, disaster prevention, and providing connectivity to internet of things devices in remote areas \cite{saeed2020cubesat}.

The capabilities of CubeSat are, however, limited by SWaP limitations, particularly for high-bandwidth communications. Furthermore, the vast potential of many CubeSat missions, is constrained not only by their communication capabilities but also by regulatory burdens in obtaining licenses for RF spectrum usage \cite{vanreusel2021launching}. The integration of CubeSats with FSO links represents a significant advancement in space-based communication systems by offering access to huge unlicensed bandwidth \cite{li2021advanced,li2022survey}. Indeed, FSO technology finds a natural synergy with the compact size and agility of CubeSats. By leveraging FSO links, CubeSats can establish high-speed, line-of-sight communication links with ground stations or other satellites, enabling real-time high-throughput data transmission without interfering with RF-based terrestrial networks. This capability is particularly valuable for applications requiring high-bandwidth communication, such as Earth observation, disaster monitoring, and remote sensing. Early studies by the Jet Propulsion Laboratory (JPL) have identified CubeSat laser communications as pivotal for a diverse range of exploratory missions \cite{mathason2019cubesat}. Significant efforts are currently underway in numerous prominent research organizations around the globe to miniaturize next-generation space-based laser communication systems, while simultaneously achieving greater data transmission rates and enhancing CubeSat capabilities \cite{li2022survey,kolev2023latest}.

Although the integration of FSO links into CubeSats opens up new opportunities for advanced capabilities, FSO-based CubeSat communication encounters several challenges that need to be carefully considered when designing link parameters and evaluating the system performance \cite{saeed2020cubesat, li2021advanced, 7553489, dresscher2019key}. In both uplink and downlink scenarios, the transmitted optical beam propagates through Earth’s atmosphere to reach the receiver (Rx), where multiple attenuations impact the amount of received power. Furthermore, the random nature of atmospheric turbulence induces signal fading, resulting in random fluctuations in received optical power. Additionally, background noise from the Sun and other stars makes it challenging to accurately detect transmitted data. Importantly, due to the narrow beamwidth of lasers, establishing and maintaining optical links becomes challenging for long-range FSO systems with moving terminals. Here, the optical Rx subsystem mounted on the CubeSat must initially acquire the pointed narrow beam laser signal from the optical transmitter (also referred to as acquisition), before establishing the link. Once successful acquisition occurs, the Rx must continuously track the angle-of-arrival (AoA) of the incoming beam to maintain link alignment (also referred to as fine beam tracking). Consequently, the successful implementation of FSO-based CubeSat communication heavily relies on the performance of pointing, acquisition, and tracking (PAT) systems. Furthermore, the power and payload constraints of CubeSats limit the size and capabilities of FSO systems that can be integrated, requiring innovative solutions to achieve efficient and reliable communication. Overcoming these challenges will be crucial for realizing the full potential of FSO-based CubeSat communications and harnessing its benefits for a wide range of space-based applications.

  
\subsection{Major Contributions and Novelty} 

In this paper, we study the problem of joint fine beam tracking and optical signal detection for a ground-to-CubeSat FSO link. To this end, we first derive a novel analytical channel model for the considered link  when the CubeSat is equipped with a converging lens and a photodetector array with an arbitrary size. Because the CubeSat is a mobile platform and moves from one point to another during the communication time interval, the coherence time of the underlying channel is reduced. Therefore, compared to fixed FSO communications, we need to perform channel estimation at shorter intervals to ensure a reliable communication. Under the circumstances, using pilot symbols for channel estimation results in a signaling overhead and spectral efficiency loss. Thus, it seems reasonable to suggest the use of blind signal detection for CubeSat-based FSO links. Consequently, we consider
a practical scenario in which the Rx has no information about the instantaneous channel coefficient and blindly detect data at the Rx. In particular, data detection is performed based on the assumption that avalanche photodiodes (APDs) are employed at the array. APDs are known for their high sensitivity and ability to detect weak optical signals, which is crucial for maintaining reliable communication in challenging conditions. We take a shot-noise limited scenario into account \cite{laserbook} (i.e., the distribution of the outputs of the APD varies as the channel states change), and thus the performance analysis of the considered FSO system is more challenging than those with positive-intrinsic-negative (PIN) photo-diodes and thermal or background noise limited scenarios.

In summary, our key contributions encompass the following aspects.
\begin{itemize}
	\item We first develop a model for the optical channel between the optical source at the Tx and the Rx detector array. This model comprehensively considers various factors, including atmospheric turbulence and attenuation, geometrical loss, CubeSat vibrations, the number and size of photodetectors in the array, and the junction width between two adjacent detectors in the array.
	\item Utilizing the derived channel model, we proceed to assess the link performance through comprehensive mathematical analysis. Our findings demonstrate that the angular instability of the CubeSat significantly influences system performance. Furthermore, we illustrate that employing an array of detectors can alleviate the adverse effects of AoA fluctuations and enhance system performance. These analytical results serve as a benchmark for determining the optimal number of APDs in the array to attain desired performance levels under varying degrees of instability without resorting to time-consuming simulations.
	\item We propose a sub-optimal blind sequence data detection method based on the generalized likelihood ratio test (GLRT) criterion, which detects data over a sequence of transmitted optical signals with length $L$ bits. Our analysis demonstrates that when $L$ is sufficiently large, the GLRT-based method can achieve performance close to that of the ideal Rx. Furthermore, although a larger value of $L$ is required to achieve performance comparable to the ideal receiver, our proposed method significantly reduces the computational load compared to the maximum likelihood (ML)-based method, which is on the order of $2^L$. This reduction in computational complexity translates to lower energy consumption and allows for the use of simpler processing units in the CubeSat payload. We also employ two combining techniques at the Rx, i.e., equal gain combining (EGC) and maximal ratio combining (MRC). However, our findings indicate that neither combining method can achieve performance comparable to the GLRT-based method.
	\item Finally, to fully leverage the benefits of utilizing a photodetector array at the Rx, we estimate the position of the beam hot spot by comparing the output of APDs. Subsequently, the instantaneous orientation of the mounted Rx with respect to the arrival direction of the received optical beam can be determined. This information can be fed back through a control message to the mechanical subsystem of the CubeSat (e.g., fast steering mirror) to correct any instantaneous orientation errors, allowing for fine-tracking of the beam while simultaneously performing data communication.
\end{itemize}

The rest of this paper is structured as follows. Section \ref{priorwork} presents the prior art. In Section \ref{II}, we detail the system model of the ground-to-CubeSat FSO link. Section \ref{III} outlines our proposed signal detection and beam tracking schemes when utilizing an APD array at the Rx. Following that, Section \ref{numerical} shows the numerical results of the proposed schemes and system performances. Finally, Section \ref{conc} provides the conclusion of this paper.

\section{Prior Art}
\label{priorwork}

Exploring the integration of FSO communication with CubeSats opens up a new frontier in research, particularly in system design, managing power efficiency and reliable data communication,  that needs to be further investigated. Prior research in this context can be categorized into two main directions: i) design and prototyping of different optical components mounted on CubeSats, as demonstrated in \cite{kolev2023latest, li2021advanced, li2022survey, janson2016nasa, cahoy2019cubesat, serra2019optical, grenfell2018pointing, riesing2022pointing}, and ii) data communication, as evidenced by \cite{samy2022ergodic, samy2023hybrid, liang2023latency, nguyen2023secrecy, le2021throughput, madoery2024novel, dabiri2024modulating}.
\subsection{Advancements in CubeSat Laser Technology}
The latest advancements in technology development and the system design of CubeSat laser communication terminals (LCTs) were demonstrated in \cite{kolev2023latest, li2021advanced, li2022survey}. For instance, NASA's optical communication and sensor demonstration (OCSD) was the first CubeSat laser payload mission in which laser terminals were mounted on a 1.5U CubeSat weighing 2.5 kg \cite{janson2016nasa}. Also, the CubeSat laser infrared crosslink (CLICK) mission showcased terminals capable of conducting full-duplex, high-data-rate crosslinks and enabling precise ranging on 3U CubeSats in LEO. The CLICK-A payload had a mass of less than 1.2 kg and fit within a 1.2U envelope \cite{cahoy2019cubesat}. Most of these LCTs were designed with the aim of reducing the cost and weight of the CubeSat payload. Consequently, it becomes feasible to utilize off-the-shelf components in ground-based optical networks, such as detector arrays, which are commercially available \cite{serra2019optical, grenfell2018pointing}. However, challenges such as AoA fluctuations arise due to CubeSat vibrations resulting from imperfectly stabilized commercial products, which often exhibit minor mechanical instabilities and less precise attitude control compared to custom-designed systems \cite{saeed2020cubesat, 7553489}. Additionally, CubeSats rely on a combination of gyroscopes, magnetometers, accelerometers, sun position sensors, Earth horizon sensors, or star trackers to determine their orientation with limited accuracy, primarily due to SWaP constraints, which induce pointing errors. In addressing these challenges, detector arrays, typically composed of APDs for long-range satellite links, play a crucial role. They enable the acceptance of the received optical beam across a wider field of view (FoV) to compensate for AoA fluctuations. Moreover, they facilitate received optical beam position sensing (AoA estimation) for fine beam tracking \cite{grenfell2018pointing, safi2021beam, riesing2022pointing}.

\subsection{Data Communication}
In the literature concerning CubeSat FSO data communication, researchers have explored various aspects to enhance the capabilities of these miniature satellite systems across different link configurations \cite{samy2022ergodic, samy2023hybrid, liang2023latency, nguyen2023secrecy, le2021throughput, madoery2024novel, dabiri2024modulating}. For instance, the study in \cite{samy2022ergodic} examined the capacity and error rate performance of an integrated transmission RF/FSO system over the uplink scenario where a high-altitude platform (HAP) is deployed as a relay between a ground station and the LEO satellite. Furthermore, the authors extended their findings from \cite{samy2022ergodic} by incorporating various channel parameters such as beam divergence loss, free space loss, and atmospheric attenuation. They analyzed the end-to-end outage probability and average symbol error probability performance in \cite{samy2023hybrid}. Moreover,  the tradeoff between satellite transmission power and network latency for different laser inter-satellite links was explored in \cite{liang2023latency}. Addressing physical layer security concerns, the authors in \cite{nguyen2023secrecy} focused on FSO-based satellite communication in LEO. Meanwhile, high-altitude ground stations as a transformative element towards an all-optical satellite megaconstellation was proposed in \cite{madoery2024novel}. The authors developed a comprehensive analytical model-based cross-layer approach for evaluating transmission control protocol (TCP) performance in the scenario of FSO-based satellite-assisted internet of vehicles \cite{le2021throughput}. More recently, the optimal design of a modulated retroreflector (MRR) laser link to establish a high-speed downlink for CubeSat FSO communication was studied in \cite{dabiri2024modulating}.
\subsection{Rationale for Our Research}
Although the idea of deploying a detector array for beam tracking has been proposed in the literature, most of them are limited to 2$\times$2 arrays, also known as quad detectors, except for a few recent works \cite{tsai2024improved}. However, these works only consider the problem of fine beam tracking with the ideal assumption of neglecting the effect of the junction width of the detector in the array (dead space), thus failing to fully capture the actual detector characteristics in the system model. Furthermore, the aforementioned works primarily focus on the channel between the Tx and receiver Rx lens. Given that CubeSat vibrations cause shifted diffraction patterns at the photodetector array \cite{7553489}, which can attenuate the received optical power, it is essential to consider an additional channel parameter to address the channel coefficient between the receiver lens and the detector array. Additionally, these prior works mainly concentrate on the problem of optical beam tracking using detector arrays, while optimal signal detection in this context when using detector arrays at the Rx has yet to be investigated. Maximizing the CubeSat's battery life is crucial for keeping replacement costs low, as a satellite's mission duration depends heavily on its battery performance. Therefore, it is vital for detection algorithms to achieve desirable performance with reduced complexity, thereby minimizing energy consumption and addressing SWaP limitations.
\section{System Model}
\label{II}
{
We consider a ground-to-CubeSat FSO link with the propagation length $Z = \frac{H_s - H_g}{\cos(\xi)}$, where $H_s$ is the sattelite altitude,  $H_g$ is the ground node altitude, and $\xi$ is the satellite’s zenith angle. The Tx and Rx nodes are positioned at coordinates ($0,0,0$) and ($0,0,Z$), respectively, in a Cartesian coordinate system ($x,y,z$) $\in \mathbb{R}^{1\times 3}$. The optical beam is propagated over the $z$-axis. Here, we assume that coarse pointing  and acquisition has been achieved, and the coarse control loop is closed, allowing the ground station to locate and track the satellite in space \cite{epple2006using, wilkerson2006concepts, kaymak2018survey}. However, the mechanical vibrations of the satellite cause fluctuations in the Rx aperture away from the boresight direction along the $z$-axis. This results in performance degradation due to fluctuations in the AoA of the received optical beam at the photodetector plane. Hence, a fine tracking stage is required to supplement the coarse stage in order to maintain the stability of the link. We employ an array of APDs at the Rx to increase its FoV to combat the adverse effects of AoA fluctuations. We also simultaneously perform fine beam tracking and data detection. The details of the proposed schemes are provided in Section \ref{III}. In the sequel, we first model the FSO channel and then present the signal model of the considered FSO setup.

	\subsection{Channel Model}
	During transmitting the optical signal by a ground node toward the CubeSat, the received signal must be collected by the CubeSat's aperture lens. The collected signal is then focused on the detector plane. Finally, the array of APDs converts the optical signal into the electrical signal.   
	Hence, as depicted in Fig. \ref{Ground-to-UAV}, the channel coefficient for the considered aerial FSO link can be divided in two different parts as
	\begin{align} \label{n2}
		h_c = h_1 h_2
	\end{align} 
	where $h_1$ represents the channel coefficient between the optical transmitter at the ground station and the receiver lens, while $h_2$ denotes the channel coefficient between the receiver lens and the array of APDs.

\begin{figure}[t]
	\begin{center}
		\includegraphics[width=3.4 in]{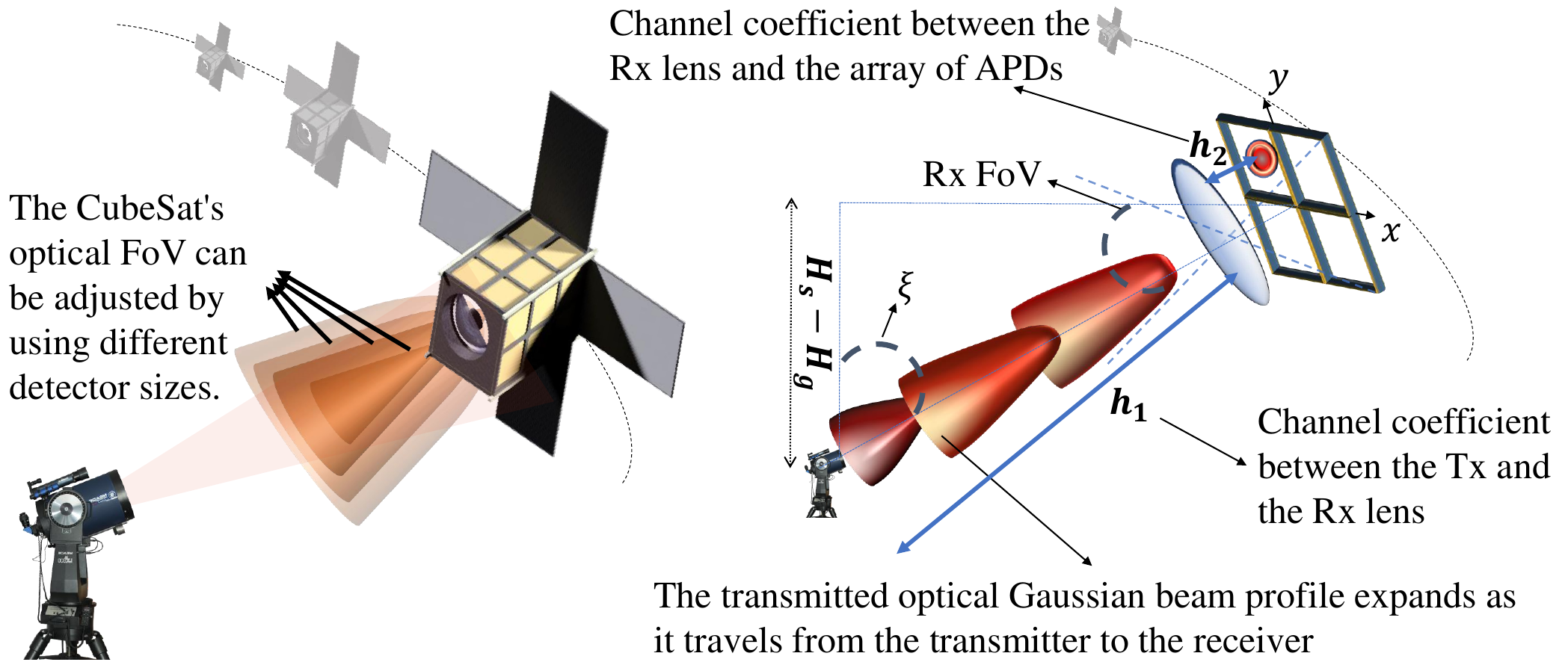}
		\caption{Graphical illustration of a ground-to-CubeSat FSO link for 6G and beyond wireless networks.}
		\label{Ground-to-UAV}
	\end{center}
\end{figure}

\subsubsection{{Channel Modelling Between the ground station and CubeSat Rx Lens}}
Let $P_{r}'=h_1 P_t$ be the collected optical power by the receiver lens where $P_t$ denotes the transmit optical power. Let $\theta_{ex}\sim \mathcal{N}(0,\sigma_{\theta_e}^2)$ and $\theta_{ey}\sim \mathcal{N}(0,\sigma_{\theta_e}^2)$ denote pointing angle errors in the directions of $x$ and $y$ axis, respectively, with $\sigma_{\theta_e}$ being the standard deviation of the angle errors \cite{dabiri2024modulating}. These errors can be induced by the limited accuracy of the CubeSat's position and orientation sensors, as well as beam wander resulting from inhomogeneities in air temperature and pressure  \cite{safi2020analytical}. The radial displacement between  the received beam center and the receiver aperture center can be represented as $d_d=\sqrt{d_{dx}^2+d_{dy}^2}$, where $d_{dx}$ and $d_{dy}$ are the displacement in the directions of $x$ and $y$ axes, respectively, and they are obtained as
	\begin{align}
		\label{po1}
		d_{dx} = Z\sin\left(\theta_{ex}\right),	~~~ \&~~~ d_{dy} = Z\sin\left(\theta_{ey}\right).
	\end{align}
	We consider a Gaussian beam at the ground station, for which the normalized spatial distribution of the received intensity at distance $Z$ is given by 
	\begin{align}
		I_r(d,Z) = \frac{2}{\pi w_z^2} \exp\left(-\frac{2(x^2 + y^2)}{w_z^2} \right),
	\end{align}
	where $d = [x, y]$ is the radial distance vector from the beam center. Also, $w_z$ is the beamwidth at distance $Z$ and can be approximated as \cite[eq. 5]{ricklin2002atmospheric}
	\begin{align}
		\label{ss1}
		w_z\simeq w_0 \sqrt{1+\left(\frac{\lambda Z}{\pi w_0^2}\right) },
	\end{align}
	where $\lambda$ is the optical wavelength and $w_0$ is the beamwidth at the output of optical transmitter.
	For lower values of $w_0$, \eqref{ss1} can be simplified as $w_z\simeq \theta_{div} Z$, where $\theta_\textrm{div}$ is the transmit optical beam divergence angle.

For a receiver aperture with the effective area of $A_r$, the effective channel coefficient due to the geometric spread with pointing error is obtained as
	\begin{align}
		\label{ss2}
		h_{p_u} &= \frac{2}{\pi w_z^2} \int \int_{p_{A_r}(x,y)}	 \nonumber \\
		&~~~\times \exp\left(-\frac{2((x-d_{dx})^2 + (y-d_{dy})^2)}{w_z^2} \right)  \textrm{d}x \textrm{d}y,
	\end{align}
	where $p_{A_r}(x,y)$ is the position of the effective aperture area in the $x-y$ plane.
Unlike terrestrial links, the laser beamwidth for satellite communications is on the order of several tens of meters. In contrast, the dimensions of the aperture in the CubeSat payload are on the order of several centimeters. Therefore, the active area of the receiver lens is usually much smaller than the received beamwidth $w_z$ in practical satellite FSO links. From this perspective, one can conclude that the optical beam maintains its plane wave nature locally at the receiver lens \cite{safi2020analytical}, and thus, \eqref{ss2} can be approximated as
	\begin{align}
		\label{ss3}
		h_{p_u} &\approx \frac{2 A_r}{\pi w_z^2} 	 
		\exp\left(-\frac{2(d_{dx}^2 + d_{dy}^2)}{w_z^2} \right) .
\end{align}

In addition to the aforementioned factors, the transmitted optical power from the ground station is affected by atmospheric turbulence-induced fading and atmospheric loss. The atmospheric attenuation is typically modeled by the Beer-Lambert law as 
	\begin{align}
		\label{ss6}
		h_{l}=\exp(-Z\zeta),
	\end{align}
where $\zeta$ is the scattering coefficient and is a function of visibility. 
	%
	%
	%
Under moderate to strong turbulence conditions, we use the Gamma-Gamma (GG) distribution to model the random variable $h_{a}$ representing turbulence conditions as
	\begin{align}
		\label{ss8}
		f_G\left(h_{a}\right) = \frac{2(\alpha\beta)^{\frac{\alpha+\beta}{2}}}{\Gamma(\alpha)\Gamma(\beta)}
		h_{a}^{^{\frac{\alpha+\beta}{2}-1}}
		k_{\alpha-\beta}  \left( 2\sqrt{\alpha\beta h_{a}}\right),
	\end{align}
where $\Gamma(\cdot)$ is the Gamma function and $k_m(\cdot)$ is the modified Bessel function of the second kind of order $m$. Also, $\alpha$ and $\beta$ are respectively the effective number of large-scale and small-scale eddies, which depend on Rytov variance $\sigma_R^2$ \cite{laserbook}. The parameter $\sigma_R^2$ is also obtained for two nodes with different altitude as \cite[p. 509]{laserbook}
	\begin{align}
		\label{sss1}
		&\sigma_R^2 = 2.25 \left( {2\pi}/{\lambda} \right)^{7/6}  \left( H_s - H_g \right)^{5/6}\sec(\xi)^{11/6} \\
		&~~~\times \int_{H_g}^{H_s} 
		C_n^2(h) \left( 1 - \frac{h - H_g}{H_s - H_g}  \right)^{5/6}  \left(\frac{h - H_g}{H_s - H_g}\right)^{5/6}  \textrm{d}h. \nonumber
	\end{align}
In \eqref{sss1}, $C_n^2(h)$ denotes the variation of the refractive index structure parameter used to characterize the atmospheric turbulence described by the Hufnagel-Valley model as
	\begin{align}
		&C_n^2(h) = 0.00594 (\mathcal{V}/27)^2 \left(10^{-5} h \right)^{10}
		\exp\left( -\frac{h}{1000} \right)  \\
		&~~~ + 2.7 \times 10^{-16}   \exp\left( -\frac{h}{1500} \right) 
		+ C_n^2(0) \exp\left( -\frac{h}{100} \right), \nonumber
	\end{align}
where $\mathcal{V}$ (in m/s) is the speed of strong wind, and $C_n^2(0)$ (in $\textrm{m}^{-2/3}$) is a strong nominal ground turbulence level. Finally, $h_1$ can be obtained as
\begin{align}
	\label{h1_final}
	h_1 = h_{p_u} h_l h_a.
\end{align}
%
%

\begin{figure}[t]
	\begin{center}
		\includegraphics[width=3.4 in]{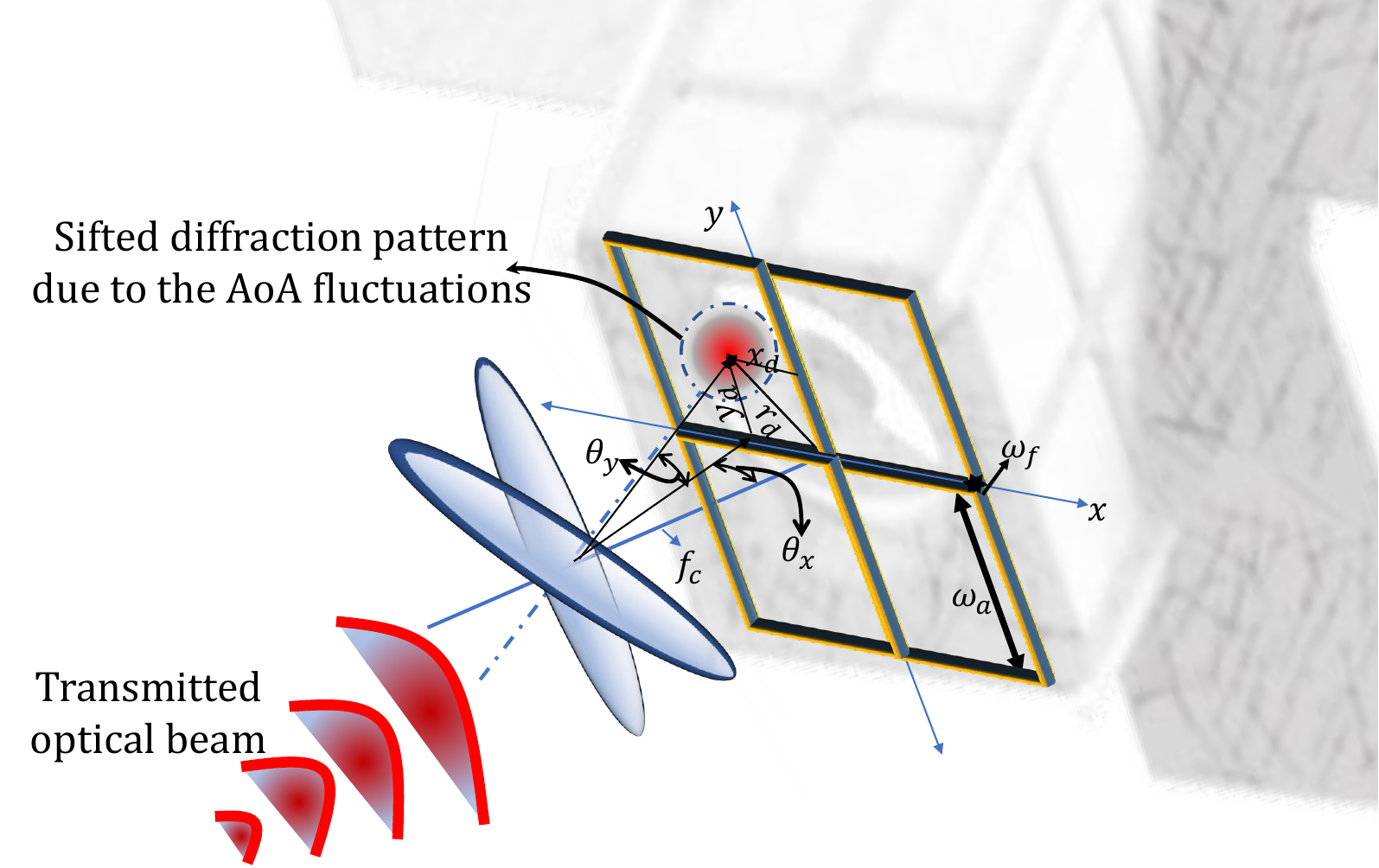}
		\caption{A schematic of the deviated received beam due to the AoA fluctuations on a $2\times 2$ APD detector.}
		\label{AoAfluctuations}
	\end{center}
\end{figure}

%

\subsubsection{Channel Modelling Between the Rx Lens and the Photodetector Array}
%
Here, we develop a mathematical model for the channel coefficient between the receiver lens and the APDs. At the Rx side, a circular lens focuses both the collected optical signal and undesired background noise, which fall within the Rx FoV, onto the photodetector array. 	As discussed, vibrations in the CubeSat cause the center of the Rx lens to deviate from the center of the received optical beam, resulting in AoA fluctuations, as graphically illustrated in Fig. \ref{AoAfluctuations}. The AoA fluctuations increase the probability that the received optical beam lies outside of the Rx FoV, significantly degrading system performance. Even when the optical signal is collected by the aperture lens, AoA fluctuations may cause the diffracted patterns to shift out of the photodetector array. This can attenuate the amount of optical power received at the Rx.

To compensate for the effect of AoA fluctuations on the link performance and to achieve a wider FoV, we utilize an array of APDs at the Rx. Specifically, we consider an array of size $N = N_a \times N_a$, where APDs are arranged next to each other in a rectangular shape, as depicted in Fig. \ref{u1}. Each APD has an active width of $w_a$, determined by a pair of indices ($i, j$). Additionally, there is a dead space between adjacent APDs, represented by the junction width $w_f$. 

%
\begin{figure}[t]
	\begin{center}
		\includegraphics[width=3 in]{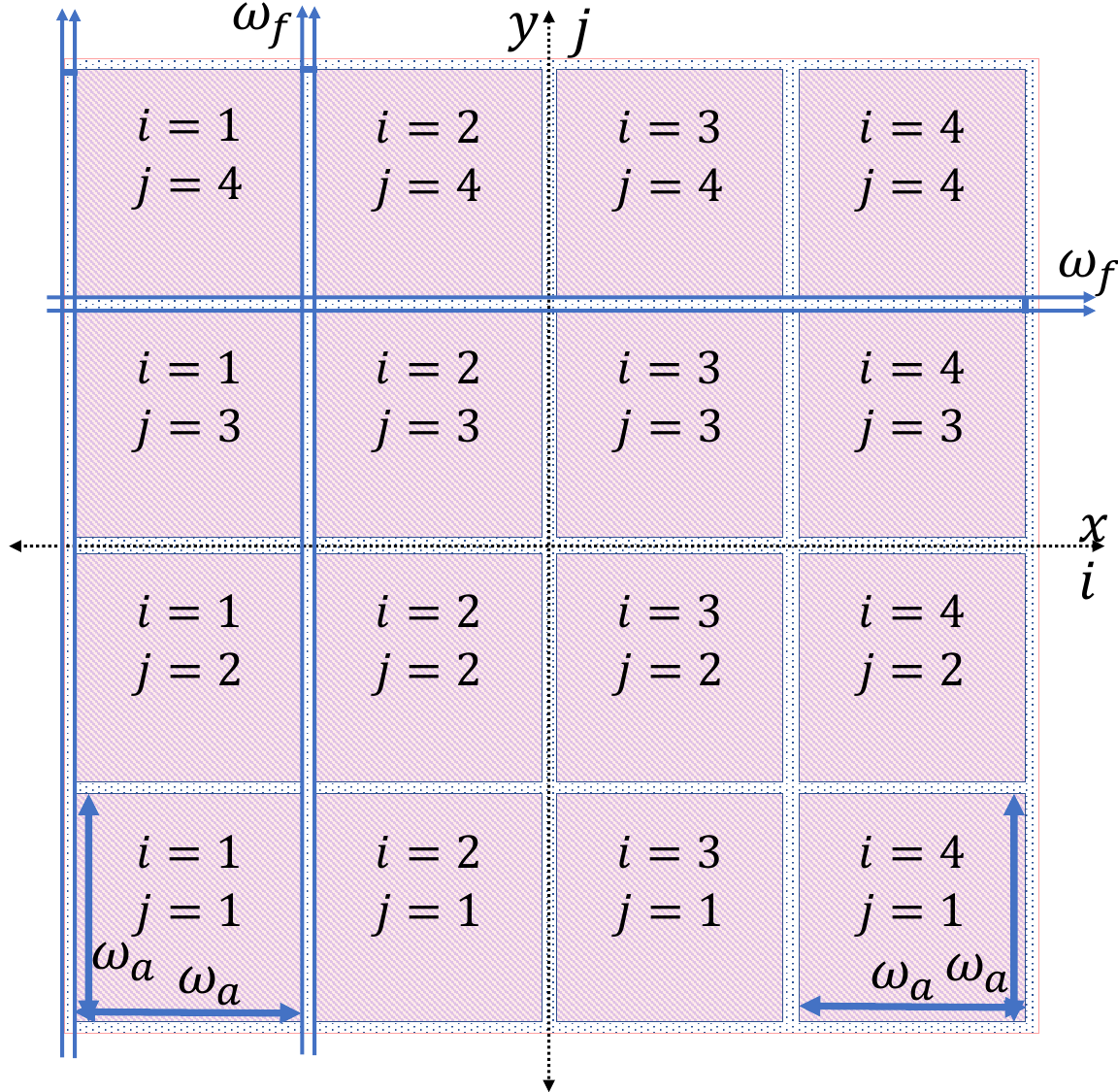}
		\caption{\textcolor{black}{The Graphical illustration of a $4\times4$ photodetector array where each photodetector is specified by $i$ and $j$ indices. Moreover, $w_f$ denotes the junction width (dead space) and $w_a$ denotes the active width of each photodetector.}}
		\label{u1}
	\end{center}
\end{figure}
%

We assume that both the Rx lens and the detector array are situated on the $x-y$ plane, while the optical beam propagates along the $z$-axis. Let's define the AoA of the signal as the incidence angle relative to the Rx axis, denoted by $\theta_r$. Consequently, $\theta_r$ can be closely approximated as (see Fig. \ref{AoAfluctuations})
	\begin{align}
		\label{AOA}
		\theta_r\simeq\sqrt{\theta_x^2+\theta_y^2}
	\end{align}
where $\theta_x$ and $\theta_y$ denote the deviations of received laser beam in $x-z$ and $y-z$ planes, respectively. At the Rx, the converging lens focuses the collected optical signals onto the area of the APD array placed at the focal plane. However, due to AoA fluctuations, the center of the optical beam at the APD area deviates, represented by $r_d = \sqrt{x_d^2+y_d^2}$, where $x_d = f_c\tan(\theta_x)\simeq f_c\theta_x$, $y_d = f_c\tan(\theta_y)\simeq f_c\theta_y$, and $f_c$ denotes the focal length. To illustrate these parameters related to the optical Rx, refer to the schematic of a $2\times 2$ APD array shown in Fig. \ref{AoAfluctuations}. The random variables (RVs) $\theta_x$ and $\theta_y$ are modeled by zero-mean Gaussian distributions with variances $\sigma_x^2$ and $\sigma_y^2$, respectively, and their joint probability density function (PDF) is obtained as \cite{j2018channel}
	\begin{align}
		\label{orientation}
		p_{\theta}\left( \theta_x,\theta_y\right) = \frac{1}{2\pi \sigma_x\sigma_y} \exp\left( -\frac{\theta_x^2}{2\sigma_x^2}  -\frac{\theta_y^2}{2\sigma_y^2}\right).
	\end{align}
We denote the Rx FoV in $x-z$ and $y-z$ planes by $\theta_{x\textrm{FoV}}$ and $\theta_{y\textrm{FoV}}$, respectively.
In this setup, we have $\theta_{x\textrm{FoV}}= \theta_{y\textrm{FoV}}= {\rm arctan}\left(\frac{N_a w_a}{f_c}\right)$. 
Therefore, the Rx FoV in the spherical coordinate system can be represented by \cite{khadjavi1968calculation}
	\begin{align}
		\label{sdfff}
		\Omega_{FoV} &= 4\arccos\left(\sqrt{\dfrac{1+u^2+v^2}{(1+u^2)(1+v^2)}}\right)
	\end{align}
	where $u = \frac{\tan\left(\theta_{x\textrm{FoV}}\right)}{2}$, and $v = \frac{\tan\left(\theta_{y\textrm{FoV}}\right)}{2}$.

The intensity of the incident optical beam on the detector array can be approximated as a two-dimensional Gaussian-shaped function \cite{gagliardi1980pointing,bashir2016optical}
	\begin{align}
		\label{n1}
		I_p(x,y) = \frac{h_1 P_t}{2\pi\sigma_I^2}
		\exp\left(-\frac{(x-f_c\theta_x)^2+(y-f_c\theta_y)^2}{2\sigma_I^2}\right)
	\end{align}
where $\sigma_I^2$ is the variance of the intensity of optical beam on the detector array. According to \eqref{n2} and \eqref{n1}, the channel coefficient corresponding to the $(i,j)$th detector can be obtained as
	\begin{align}\label{n3}
		h_{ij} = h_1 h_{2,ij}
	\end{align}
where 
	\begin{align}
		\label{n4}
		h_{2,ij} &= \int_{(i-1-\frac{N_a}{2})w_a+w_f/2}^{(i-\frac{N_a}{2})w_a-w_f/2}
		\int_{(j-1-\frac{N_a}{2})w_a+w_f/2}^{(j-\frac{N_a}{2})w_a-w_f/2} \frac{1}{2\pi\sigma_I^2} \nonumber\\
		&~~~\times\exp\left(-\frac{(x-f_c\theta_x)^2+(y-f_c\theta_y)^2}{2\sigma_I^2}\right) \textrm{d}x\,\textrm{d}y \nonumber \\
		& =\Bigg[ Q\left(\frac{(i-1-N_a/2)w_a+w_f/2-f_c\theta_x}{\sigma_I} \right) \nonumber \\
		&~~~-Q\left(\frac{(i-N_a/2)w_a-w_f/2-f_c\theta_x}{\sigma_I} \right)\Bigg] \nonumber \\
		&~~~\times \Bigg[ Q\left(\frac{(j-1-N_a/2)w_a+w_f/2-f_c\theta_y}{\sigma_I} \right) \nonumber \\
		&~~~-Q\left(\frac{(j-N_a/2)w_a-w_f/2-f_c\theta_x}{\sigma_I} \right)\Bigg].
\end{align}

\subsection{Received Signal Model}	
Due to its relatively low implementation complexity, most current commercial FSO systems utilize intensity modulation with direct detection (IM/DD) based on on-off keying (OOK). In practical FSO links, the mean number of absorbed photons is typically sufficiently large, allowing the distribution of the number of APD output electrons to be well approximated by a Gaussian distribution \cite{davidson1988gaussian}.
Accordingly, the photo-current corresponding to the $k$th bit interval on  the $(i,j)$th detector can be expressed as
\begin{equation}
\label{f3}
r_{ij}[k] =  h_{ij} \mu s[k] + n_{ij}[k]
\end{equation}
where $i$ and $j\in\{1,...,N_a\}$, $\mu= \frac{e G \eta}{h_p \nu}$, $e$ denotes the charge of electron, $G$ is the average APD gain, $\eta$ denotes the APD quantum efficiency, $h_p$ denotes the Planck\textquotesingle s constant, and $\nu$ is the optical frequency. Moreover in \eqref{f3}, $s[k]$ and $n_{ij}[k]$, respectively, denote the transmitted symbol with optical power $P_t$, and the photo-current noise of the $(i,j)$th detector that is an additive Gaussian noise having zero-mean and variance $\sigma_{ij,k}^2$ given by
\begin{align}
\label{f4}
\sigma_{ij,k}^2 = \sigma_s^2 h_{ij} s[k] + \sigma_0^2, 
\end{align}
where $\sigma_s^2$ is the variance of the shot noise\footnote{This noise is attributed to the quantum nature of light, where the number of photons emitted by a coherent optical source in a given time is never constant. Known as quantum noise or photon noise, this phenomenon arises from the random arrival rate of photons from the data-carrying optical source, acting as a shot noise present in all photodetectors \cite{ghassemlooy2012optical}.} due to transmitted signal, and $\sigma_0^2 = \sigma_{b}^{2} + \sigma_{th}^{2}$ is the total noise variance due to the variance of background radiation, $\sigma_{b}^{2}$, and receiver thermal noise, $\sigma_{th}^{2}$. We have $\sigma_{b}^{2} = 2 e G F \mu B P_b$, where $F$ denotes the APD excess noise factor, and $B$ is the bandwidth of the receiver low-pass filter (in Hz). Furthermore, $\sigma_{th}^{2}=\frac{4K_{B}T_{r}B}{R_{l}}$, where $ K_{B} $ is Boltzmann constant, $ T_{r} $ is the receiver equivalent temperature in Kelvin, and $ R_{l} $ is the load resistance. 
The background power $P_b$ is a function of the photodetector area and can be obtained as \cite{j2018channel, 7553489}
\begin{align}
\label{xf1}
P_b = N_b(\lambda)\, B_o\, \Omega_{FoV}\, A_a
\end{align}
where $N_b(\lambda)$ is the spectral radiance of the background radiations at wavelength $\lambda$ (in Watts/${\rm cm}^2$-$\micro$m-srad), $B_o$ is the bandwidth of the optical filter at the Rx (in $\micro$m), and $A_a = \pi r_a^2$ is the lens area (in ${\rm cm}^2$).

\section{Data Detection and Spatial Beam Tracking}
\label{III}
%

We assume that the transmitted optical signal is captured during an observation window consisting of $L$ bits. Therefore, the received signal vector $\underline{r}_{ij} = \left[{r_{ij}[1],\ldots,r_{ij}[L]}\right]$ at the $(i,j)$th APD is related to the $L$ transmitted signal vector $\underline{s} = \left[{s[1],s[2],\ldots,s[L]}\right]$.  We also assume slow fading channel, i.e., the channel remains constant during the observation window \cite{7553489}.

In this paper, we address a practical scenario where channel state information (CSI) and the beam center position (or equally beam AoA) are unknown at the Rx side. Our objective is to detect OOK signals without the need for pilot symbols and to determine the position of the incident optical beam on the array of APDs. To achieve this, we propose an efficient data-aided channel estimation method without requiring pilot symbols. We then detect the sequence of OOK signals and perform spatial beam tracking. We evaluate the performance of our sub-optimal data detection method in terms of bit error rate BER, considering the ideal scenario where perfect CSI and exact beam position are available at the Rx as a benchmark. Further details of these methods are presented in the following sections.

\subsection{Ideal Data Detection and Spatial Beam Tracking}
In this subsection, we first study the ideal data detection under the assumption of perfect knowledge of the channel coefficient $h_1$, and the beam angular deviations $\theta_x$ and $\theta_y$, as a benchmark. 
\subsubsection{Ideal Data Detection}

For the ideal detection method, we assume that the Rx perfectly knows the instantaneous channel coefficient $h_1$. We also assume that the values of $\theta_x$ and $\theta_y$ are available with high accuracy (i.e., no need to spatial beam tracking). Hence, the position of the center of optical beam spot is available, which leads to the perfect knowledge of the channel coefficient $h_{2,ij}$.

Under this ideal scenario, the ideal detection method based on the ML criterion can be performed symbol-by-symbol as
\begin{align}
\label{f1}
\hat{s}[k] = \underset{s[k]\in\{0,1\}}{\operatorname{\text{~arg~max~}}}~
\prod_{i=1}^{N_a} \prod_{j=1}^{N_a} p(r_{ij}[k]|h_{ij};s[k])
\end{align} 
where $p(\cdot)$ denotes the probability of an event.
Accordingly, from \eqref{f3}, \eqref{f4}, and \eqref{f1}, the ideal data detection method can be obtained as

\begin{gather}
\label{f2}
 \sum_{i=1}^{N_a}\sum_{j=1}^{N_a}  
\frac{-2\sigma_0^2\mu h_{ij}r_{ij}[k] + \sigma_0^2\mu^2 h_{ij}^2     - \sigma_s^2 h_{ij}r_{ij}[k]^2 }
{\sigma_0^2\sigma_s^2 h_{ij}  + \sigma_0^4}  ~  
\substack{\hat{s}[k]=0 \\ >\\< \\ \hat{s}[k]=1} 
~\sum_{i=1}^{N_a}\sum_{j=1}^{N_a}  \ln\left( \frac{\sigma_0^2}{\sigma_s^2 h_{ij}  + \sigma_0^2} \right).
\end{gather}
Moreover, the BER of the ideal detection method is obtained as
\begin{gather}
\label{e2}
\mathbb{P}^\textrm{ML}_e = \int_0^\infty  \int_0^\infty  \int_0^\infty \mathbb{P}^\textrm{ML}_{e|h_1,\theta_x,\theta_y}  
\times f_{h_1}(h_1) f_{\theta_x}(\theta_x) f_{\theta_y}(\theta_y) \textrm{d}h_1 \textrm{d}\theta_x \textrm{d}\theta_y
\end{gather}
where
\begin{align}
\label{e3}
\mathbb{P}^\textrm{ML}_{e|h_1,\theta_x,\theta_y} = p_0 \mathbb{P}^\textrm{ML}_{e|01} + p_1 \mathbb{P}^\textrm{ML}_{e|10}.
\end{align}
In \eqref{e3}, $p_0$ and $p_1$ denote the probability of transmitting bit ``0'' and ``1'', respectively, and it is assumed that $p_0=p_1= \sfrac{1}{2} $. Also, $\mathbb{P}_{e|01}$ and $\mathbb{P}_{e|10}$ denote the BER conditioned on $s[k]=0$ and $s[k]=1$, respectively, and they can be obtained as  
\begin{align}
\label{k1}
\mathbb{P}^\textrm{ML}_{e|10} &= \textrm{Prob}\Bigg\{\sum_{i=1}^{N_a}\sum_{j=1}^{N_a} \bigg[  
-2\sigma_0^2\mu h_{ij}(\mu h_{ij} + n_{ij}[k])    \\
&~~~+ \sigma_0^2\mu^2 h_{ij}^2 - \sigma_s^2 h_{ij}(\mu^2 h_{ij}^2 + 2\mu h_{ij} n_{ij}[k] + n_{ij}^2[k]) \nonumber \\
&~~~- (\sigma_0^2\sigma_s^2 h_{ij}  + \sigma_0^4)\ln\left( \frac{\sigma_0^2}{\sigma_s^2 h_{ij}  + \sigma_0^2} \right)   \bigg] 
> 0 \Bigg\}  \nonumber\\
& \simeq \textrm{Prob}\Bigg\{  \sum_{i=1}^{N_a}\sum_{j=1}^{N_a} \bigg[  
n_{ij}[k]     
+ \frac{  \mu h_{ij}}{2} 
+ \frac{(\sigma_0^2\sigma_s^2 h_{ij}  + \sigma_0^4)}{2\mu h_{ij} (\sigma_0^2  + \sigma_s^2 h_{ij})} \nonumber \\
&~~~\times\ln\left( \frac{\sigma_0^2}{\sigma_s^2 h_{ij}  + \sigma_0^2} \right)   \bigg] 
< 0 \Bigg\} \nonumber
\end{align} 
and
\begin{align}
\label{k2}
\mathbb{P}^\textrm{ML}_{e|01}&= \textrm{Prob}\Bigg\{ \sum_{i=1}^{N_a}\sum_{j=1}^{N_a} \bigg[  
\frac{-2\sigma_0^2\mu h_{ij}n_{ij}[k] + \sigma_0^2\mu^2 h_{ij}^2  }
{\sigma_0^2\sigma_s^2 h_{ij}  + \sigma_0^4}  \\
&~~~-\frac{\sigma_s^2 h_{ij}n_{ij}[k]^2 }
{\sigma_0^2\sigma_s^2 h_{ij}  + \sigma_0^4}
- \ln\left( \frac{\sigma_0^2}{\sigma_s^2 h_{ij}  + \sigma_0^2} \right)   \bigg] 
< 0 \Bigg\} \nonumber \\
& \simeq \textrm{Prob}\Bigg\{\sum_{i=1}^{N_a}\sum_{j=1}^{N_a} \bigg[  
{n_{ij}[k] - \frac{\mu h_{ij}}{2}  } 
+ \frac{\sigma_0^2\sigma_s^2 h_{ij}  + \sigma_0^4}{2\sigma_0^2\mu h_{ij}} \nonumber \\
&~~~\times\ln\left( \frac{\sigma_0^2}{\sigma_s^2 h_{ij}  + \sigma_0^2} \right)   \bigg] 
>0 \Bigg\}.\nonumber
\end{align}
The approximations utilized in \eqref{k1} and \eqref{k2} stem from neglecting the impact of $n^2_{ij}[k]$ in comparison to $n_{ij}[k]$. As demonstrated in the numerical results, this approximation proves to be reasonable, particularly in high signal-to-noise ratio (SNR) regimes.
Furthermore, considering that the error probability expression in \eqref{k1} is formulated for the scenario of transmitting bit ``1,'' it becomes evident from \eqref{f4} that the term $\sum_{i=1}^{N_a}\sum_{j=1}^{N_a} n_{ij}[k]$ constitutes a zero-mean Gaussian noise with a variance of $N_a^2\sigma_0^2+\sigma_s^2\sum_{i=1}^{N_a}\sum_{j=1}^{N_a}h_{ij}$. Likewise, the term $\sum_{i=1}^{N_a}\sum_{j=1}^{N_a} n_{ij}[k]$ in \eqref{k2} corresponds to Gaussian noise with a mean of zero and a variance of $N_a^2\sigma_0^2$, as it is derived for the transmission of bit ``0.'' Therefore, we have
\begin{gather}
\label{k3}
\mathbb{P}^\textrm{ML}_{e|10} \simeq  
 Q\left( \frac{ \sum_{i=1}^{N_a}\sum_{j=1}^{N_a}  \left[ 
	\frac{\mu h_{ij}}{2} 
	+ \frac{(\sigma_0^2\sigma_s^2 h_{ij}  + \sigma_0^4)}{2\mu h_{ij} (\sigma_0^2  + \sigma_s^2 h_{ij})} 
	\ln\left( \frac{\sigma_0^2}{\sigma_s^2 h_{ij}  + \sigma_0^2} \right) \right]}
{N_a^2\sigma_0^2+\sigma_s^2\sum_{i=1}^{N_a}\sum_{j=1}^{N_a}h_{ij}}  
\right) 
\end{gather} 
and 
\begin{gather}
\label{k4}
\mathbb{P}^\textrm{ML}_{e|01} \simeq  
Q\left( \frac{ \sum_{i=1}^{N_a}\sum_{j=1}^{N_a}  \left[ 
	\frac{\mu h_{ij}}{2} 
	- \frac{\sigma_0^2\sigma_s^2 h_{ij}  + \sigma_0^4}{2\sigma_0^2\mu h_{ij}}
	\ln\left( \frac{\sigma_0^2}{\sigma_s^2 h_{ij}  + \sigma_0^2} \right) \right]}
{N_a^2\sigma_0^2}  
\right).  
\end{gather}

\subsubsection{Ideal Spatial Tracking}
In this scenario, we assume the case that the Rx only knows the instantaneous channel coefficients $h_1$ and the vector of transmitted signal $\underline{s}$ and needs to estimate position of the beam centre on the detector array. To do so, one can obtain the number of bits ``1'' in the observation window of length $L$ as $m = \sum\limits_{k = 1}^{{L}} {s[k]}$.
Under this ideal scenario, the AoA of the received beam can be obtained  based on the ML criterion as
\begin{align}
\label{f5}
\hat{\theta}_x,\hat{\theta}_y = \underset{\theta_x,\theta_y}{\operatorname{\text{~arg~max~}}}~
\prod_{i=1}^{N_a} \prod_{j=1}^{N_a} p(r''_{ij}|h_{ij},m;\theta_x,\theta_y)
\end{align} 
where 
\begin{align}
\label{f6}
r''_{ij} &=\frac 1 m\sum_{k=1}^L r_{ij}[k]s[k] =    \mu h_{ij} + n_{ij}''
\end{align}
and $n''$ is a zero-mean Gaussian noise with variance 
\begin{align}
\label{f7}
\sigma_{n_{ij}''}^2 = \frac{\sigma_s^2 h_{ij}  +  \sigma_0^2}{m}.
\end{align}
Accordingly, from \eqref{n4}, \eqref{f6}, and \eqref{f7}, the ideal spatial tracking method of \eqref{f5} can be simplified as
\begin{align}
\label{f8}
\hat{\theta}_x,\hat{\theta}_y &= \underset{\theta_x,\theta_y}{\operatorname{\text{arg~min~}}}
\sum_{i=1}^{N_a} \sum_{j=1}^{N_a} \Bigg[ \frac{1}{m}\ln\left(\frac{\sigma_s^2 h_1h_{2,ij}(\theta_x,\theta_y)  +  \sigma_0^2}{m}\right) \nonumber\\
&~~~~~~+ \frac{|r_{ij}'' - \mu h_1h_{2,ij}(\theta_x,\theta_y) |^2}{\sigma_s^2 h_1 h_{2,ij}(\theta_x,\theta_y)  +  \sigma_0^2} \Bigg].
\end{align}

\subsection{Optimal Data Detection Method Without Knowing CSI}
Here, we consider a general condition where the Rx lacks any information regarding the instantaneous channel coefficients. Consequently, the ML detector utilizes the outputs of all APDs in the array to estimate the transmitted signal vector as 
\begin{align}
\label{f9}
\hat{\underline{s}} = \underset{\underline{s}}{\operatorname{\text{~arg~max~}}}~
\prod_{i=1}^{N_a} \prod_{j=1}^{N_a} p(\underline{r}_{ij}|\underline{s})
\end{align}
where
\begin{align}
\label{f10}
p(\underline{r}_{ij}|\underline{s}) = \prod_{k=1}^L p(r_{ij}[k]|s[k])
\end{align}
and
\begin{align}
\label{f12}
p(r_{ij}[k]|s[k]) &= \int_0^\infty \int_0^\infty \int_0^\infty p(r_{ij}[k]|s[k],h_1, h_{2,ij}(\theta_x,\theta_y)) \nonumber \\
&~~~\times f_{h_1}(h_1) f_{\theta_x}(\theta_x) f_{\theta_y}(\theta_y)
\textrm{d}h_{1} \textrm{d}\theta_x \textrm{d}\theta_y.
\end{align}
As we realize from \eqref{f9}, across an observation window of length $L$, the ML-based method can detect the transmitted signal vector $\underline{s}$. Specifically, data detection entails searching over $2^L$ potential states of the metric $\eqref{f9}$, necessitating the computation of a three-level integral. Evidently,this approach incurs relatively high computational complexity, which contradicts the limited energy consumption constraints of CubeSats. In the sequel, we seek to find  sub-optimal detection methods having lower computational complexity.
 
\subsection{Equal Gain Combining Method}
To reduce the computational complexity of the ML-based detection method, one can employ the  EGC technique at the Rx. More precisely, this approach  obviates the necessity  to decide on a particular APD by combining the output of $N_a\times N_a$ APDs with the same factor.  In theory, the output photo-current of the EGC method for an $N_a\times N_a$ APD array is similar to that of a single APD with a large active area $A_s = N_a\times N_a \times {w_a}^2$. However, unlike their terrestrial counterparts that utilize very small APDs, for the ground-to-CubeSat FSO systems, the area of the single APD must be large enough to overcome the effect of AoA fluctuations. Nevertheless, as the size of an APD increases, its electrical bandwidth and its quantum efficiency  decrease. Given the importance of maximizing detector sensitivity to minimize the required laser power for a specified link margin, rather than employing a single large APD, we opt for an array of APDs and implement the EGC technique to aggregate the total photo-current generated by the array.   

Accordingly, we denote by $r[k]$ the total photo-current generated by all APDs in the detector array, which can be expressed as
\begin{align}
	\label{xd1}
	r[k] = \sum_{i=1}^{N_a} \sum_{j=1}^{N_a} {r_{ij}[k]}  =  \mu s[k] h_t + n[k]
\end{align}
where
\begin{align}
	\label{t1}
	h_t = \sum_{i=1}^{N_a} \sum_{j=1}^{N_a} h_{ij}
\end{align}
and $n[k] = \sum_{i=1}^{N_a} \sum_{j=1}^{N_a} {n_{ij}[k]}$ is an additive Gaussian noise with zero-mean and variance
$\sigma_{k}^2 = \sigma_s^2 h_t s[k] + N_a^2\sigma_0^2$.
When $h_t$ is known at the Rx side, transmitted data can be detected symbol-by-symbol as
\begin{align}
\label{k5}
r[k] \substack{\hat{s}[k]=1 \\ >\\< \\ \hat{s}[k]=0} h_\textrm{th,EGC}
\end{align}
where $h_\textrm{th,EGC}$ is the detection threshold for EGC method. From \eqref{xd1}, the BER of the EGC method conditioned on $h_t$ can be obtained as
\begin{align}
\label{k6}
\mathbb{P}^\textrm{EGC}_{e|h_t} = \frac 12 Q\left(\frac{\mu h_t - h_\textrm{th,EGC}}{\sqrt{N_a^2\sigma_0^2 + \sigma_s^2h_t}}\right) 
+\frac 12 Q\left(\frac{h_\textrm{th,EGC}}{N_a\sigma_0}\right).
\end{align}
The optimal value for $h_\textrm{th,EGC}$ is obtained by differentiating \eqref{k6} with respect to $h_\textrm{th,EGC}$ and setting the results equal to zero. By doing so, the optimal value for $h_\textrm{th,EGC}$ is obtained as
\begin{align}
\label{k7}
h_\textrm{th,EGC}=&  
\frac{ - \mu h_t N_a^2\sigma_0^2}    {\sigma_s^2h_t} +
\frac{ \sqrt{(N_a^2\sigma_0^2 + \sigma_s^2h_t)N_a^2\sigma_0^2} }    {\sigma_s^2h_t}  \\
& \times \sqrt{\mu^2h_t^2 +2\sigma_s^2h_t\ln {\displaystyle{\left(\frac{N_a^2\sigma_0^2 + \sigma_s^2h_t}{N_a^2\sigma_0^2}     \right)}}}.\nonumber
\end{align}
However, as we will observe in the numerical results section, despite its simplicity, the EGC method fails to achieve ideal performance, even with perfect channel estimation. Specifically, since the receiver in this method incorporates all captured background noise with equal gain in the detection process, it becomes more susceptible to variations in background noise levels\footnote{{In this setup, each detector in the array adds background noise to the total noise power. Additionally, the combining algorithm does not provide an optimal detection threshold to effectively mitigate the noise introduced by each detector.}}.  Moreover, by summing the output of the APDs in the array, performing the fine beam tracking method is not directly possible. This process requires additional hardware, which may not always be available due to SWaP limitations.

In the sequel, we propose a sub-optimal method capable of data detection and beam tracking, achieving performance close to the ideal receiver without requiring knowledge of the channel. Simultaneously, this method aims to reduce the computational load associated with the optimal ML detection method.
\subsection{Sub-Optimal Detection Method and Spatial Beam Tracking}
In this subsection, we propose a sub-optimal detection method to reduce the computational complexity of the optimal ML detection method proposed in \eqref{f9}. We also aim to achieve performance close to the ideal case. 
To this end, we collect the received vector $r_{ij}$ of all APDs over an observation window of length $L$. We then use the GLRT principle to  estimate jointly the instantaneous channel coefficient of each APD $h_{ij}$, and detect the transmit data sequence. The proposed GLRT-based detection method uses the past detected data to improve the accuracy of the channel estimation.
For the considered system model, the transmitted vector $\underline{s}$ and the instantaneous channel coefficient $h_{ij}$ can be jointly estimated based on the GLRT criterion as \cite{song2014robust}
\begin{align}
\label{d1}
\hat{\underline{s}},\hat{h}_{ij} &= 
\underset{\underline{s},h_{ij}\textrm{s}}{\operatorname{\text{arg~max~}}} 
\prod_{i=1}^{N_a} \prod_{j=1}^{N_a} p(\underline{r}_{ij}|\underline{s},h_{ij}) \\
& = 
\underset{\underline{s},h_{ij}}{\operatorname{\text{arg~max~}}} 
\prod_{i=1}^{N_a} \prod_{j=1}^{N_a} \prod_{k=1}^L 
\frac{1}{\sqrt{2\pi(\sigma_s^2 h_{ij} s[k]+\sigma_0^2)}} \nonumber  \\
&~~~\times\exp\left(-\frac{|r_{ij}[k]-\mu h_{ij}s[k]|^2}
{2(\sigma_s^2 h_{ij} s[k]+\sigma_0^2)}\right).\nonumber
\end{align}
From \eqref{d1}, when $\underline{s}$ is known at the Rx side, the RV $h_{ij}$ for the $(i,j)$th APD can be estimated as
\begin{align}
\label{d2}
\hat{h}_{ij}&= 
\underset{h_{ij}}{\operatorname{\text{arg~min~}}}   
\underbrace{\sum_{k''\in\mathcal{K}''}\left(\ln(\sigma_0^2)  +  \frac{|r_{ij}[k'']|^2}{\sigma_0^2}\right)}_
\text{A1} 
 \\
&+\underbrace{\sum_{k'\in\mathcal{K}'}
\left( \ln(\sigma_s^2 h_{ij} +\sigma_0^2) 
+   \frac{|r_{ij}[k']-\mu h_{ij}|^2}       {\sigma_s^2 h_{ij} +\sigma_0^2} \right)}_
\textrm{A2} \nonumber
\end{align}
where $\mathcal{K}'$ and $\mathcal{K}''$ are the subsets of the set $\mathcal{K}=\{k\}$ where $ k \in \{1,\ldots,L\}$ for which $s[k]=1$, and $s[k]=0$, respectively. As evident from \eqref{d2}, the term $A1$ is independent of the RV $h_{ij}$. Hence, \eqref{d2} can be simplified as
\begin{align}
\label{d3}
\hat{h}_{ij} &= 
\underset{h_{ij}}{\operatorname{\text{arg~min~}}} 
\sum_{k'\in\mathcal{K}'}  
\left( \ln(\sigma_s^2 h_{ij} +\sigma_0^2) 
	+   \frac{|r_{ij}[k']-\mu h_{ij}|^2}       {\sigma_s^2 h_{ij} +\sigma_0^2} \right).
\end{align}
For any given $\underline{s}$, the optimal value of $h_{ij}$ is obtained by differentiating  \eqref{d3} with respect to $h_{ij}$ and setting the resulting expression to zero. By doing so, we have
\begin{align}
\label{d4}
&h_{ij}^\textrm{opt} = \frac{-\sigma_s^2}{2  \mu^2}  -  \frac{\sigma_0^2}{\sigma_s^2} + 
\sqrt{\frac{\sigma_s^4}{4 \mu^4}  
	+  \frac{\sigma_0^4}{\sigma_s^4}   
	+\mathbb{R}_{ij}                             }. 
\end{align} 
where $\mathbb{R}_{ij}=\frac{\sum_{k=1}^L(2\mu\sigma_0^2r_{ij}[k]s[k]  + \sigma_s^2 r_{ij}[k]^2s[k])}{m \mu^2\sigma_s^2}$.
Now, by substituting \eqref{d4} in \eqref{d1} and after some manipulations, the GLRT-based method for data detection \eqref{d1} is simplified as \eqref{d5}.
%
%
\begin{figure*}[t]
	\normalsize
	%
	\begin{align}
	\label{d5}
	\hat{\underline{s}} &
	= \underset{\underline{s}}{\operatorname{\text{arg~min~}}} 
	N_a^2(L-m)\ln( \sigma_0^2)
	+   N_a^2 \sum_{k''\in\mathcal{K}''} \frac{|r_{ij}[k'']|^2} { \sigma_0^2}
	+ m\sum_{i=1}^{N_a} \sum_{j=1}^{N_a}  
	\ln\left( \sqrt{\frac{\sigma_s^8}{4 \mu^4}  +  \sigma_0^4   	+\sigma_s^4 \mathbb{R}_{ij}}
	-\frac{\sigma_s^4}{2  \mu^2}\right) \\  
	&~~~~~~~~~~~~~~ +\sum_{i=1}^{N_a} \sum_{j=1}^{N_a} 
	\frac{1}{ \sqrt{\frac{\sigma_s^8}{4 \mu^4}  +  \sigma_0^4   	+\sigma_s^4 \mathbb{R}_{ij}}
		-\frac{\sigma_s^4}{2  \mu^2}}
	\sum_{k'\in\mathcal{K}'} 
	\left(r_{ij}[k']+\frac{\sigma_s^2}{2  \mu}  +  \frac{\mu\sigma_0^2}{\sigma_s^2}- 
	\sqrt{\frac{\sigma_s^4}{4 \mu^2}  
		+  \frac{\mu\sigma_0^4}{\sigma_s^4}   +\mu\mathbb{R}_{ij}}\right)^2.         \nonumber    
	\end{align}
	\hrulefill
\end{figure*}
It is noteworthy that unlike the metric in \eqref{f9}, which necessitates the calculation of a three-level integral, the detection metric in \eqref{d5} involves straightforward additions and multiplications, and does not rely on channel knowledge. The computational complexity of the metric in \eqref{d5} is $\mathcal{O}(LN_a^2)$. Moreover, to detect the transmitted vector $\underline{s}$ using the proposed detection method in \eqref{d5}, one needs to search among $2^L$ possible received sequences and select the sequence that minimizes the metric. Hence, the total computational complexity of the proposed detection method is $\mathcal{O}(LN_a^2\times2^L)$.

We note that for an all-zero transmitted sequence, i.e., when  $m=0$ \& $\mathbb{R}_{ij}=\frac 00$, our proposed method is unable to  estimate properly the channel, and subsequently, perform data detection. For an observation window of length $L$, the occurrence probability of an all-zero sequence is equal to $\frac{1}{2^L}$. Hence, the value of $L$ must be large enough to ensure that the
occurrence probability of the all-zero sequence is lower than the desired BER. As we will show in the numerical result section, 
the proposed GLRT-based method achieves an acceptable performance compared with the ideal receiver when $L>20$. For such values of $L$, detection is done by searching over  $2^{20}=1048576$ possible states of metric \eqref{d5}. To avoid this exhaustive search and to reduce further the Rx complexity, we propose a sub-optimal method for implementing $\eqref{d5}$,  which reduces the search space  to $L$. Particularly, by considering the nature of the OOK modulation, one can reasonably assume that the maximum received signal levels are due to the transmitted bit ``1'', and the minimum received signal levels are due to the transmitted bits ``0''.  Hence, the $m$ maximum received signal levels during the observation interval $L$ can be selected as bit ``1'', and the remainder can be selected as bit ``0''. Since the value of RV $m$  varies from   $1$ to $L$, the
Rx searches among $L$ possible received sequences (i.e., sequences with which the value of $m$ is equal to $1, 2, \ldots, L$) to find the sequence that minimizes $\eqref{d5}$. \textcolor{black}{This way, the search space for finding $\hat{\underline{s}}$ is reduced from $2^L$ to $L$, and thus, the computational complexity is obtained as $\mathcal{O}(L^2N^2_a)$.} In the numerical result section, we show that searching over this sub-space  achieves a performance close to the case that the detection is done over all possible states. 

Finally, to perform fine beam tracking by determining the position of the incident optical baeam on the APD array, we resort to the following proposition.

{\bf Proposition:} {\it Suppose the center of received optical beam is located in the ($i',j'$)th APD in the array, i.e., we have
$\Big\{(i'-1-\frac{N_a}{2})w_a+\frac{w_f}{2} <\theta_x f_c< (i'-\frac{N_a}{2})w_a-\frac{w_f}{2}\Big\}$, and 
$\Big\{(j'-1-\frac{N_a}{2})w_a+\frac{w_f}{2} <\theta_y f_c< (j'-\frac{N_a}{2})w_a-\frac{w_f}{2}\Big\}$. In this case, the channel coefficient  corresponding to that APD, $h_{2,i'j'}$, is greater than the other channel coefficients corresponding to the other APDs in the array, i.e.,  $h_{2,i'j'} \geq h_{2,ij}$, where $(i,j) \in\{1,...,N_a\}$.} \\
\begin{IEEEproof}
Let rewrite \eqref{n4} as $h_{2,ij}=h_{2,ij}^1 h_{2,ij}^2$, where 
\begin{align}
\label{x1}
&h_{2,ij}^1=\Bigg[ Q\left(\frac{(i-1-N_a/2)w_a+w_f/2-f_c\theta_x}{\sigma_I} \right) \nonumber \\
&~~~~~-Q\left(\frac{(i-N_a/2)w_a-w_f/2-f_c\theta_x}{\sigma_I} \right)\Bigg]  \\
\label{x2}
&h_{2,ij}^2=\Bigg[  Q\left(\frac{(j-1-N_a/2)w_a+w_f/2-f_c\theta_y}{\sigma_I} \right) \nonumber\\
&~~~~-Q\left(\frac{(j-N_a/2)w_a-w_f/2-f_c\theta_x}{\sigma_I} \right)\Bigg]. 
\end{align}
From \eqref{x1} and \eqref{x2}, it can be easily realized that $\Big\{(i'-1-\frac{N_a}{2})w_a+\frac{w_f}{2} <\theta_x f_c< (i'-\frac{N_a}{2})w_a-\frac{w_f}{2}\Big\}$, implies that $h^1_{2,i'j'}\geq h^1_{2,ij}$,  and, similarly,  $\Big\{(j'-1-\frac{N_a}{2})w_a+\frac{w_f}{2} <\theta_y f_c< (j'-\frac{N_a}{2})w_a-\frac{w_f}{2}\Big\}$, implies that $h^2_{2,i'j'}\geq h^2_{2,ij}$. Thereby, $h^1_{2,i'j'}\geq h^1_{2,ij}$ and $h^2_{2,i'j'}\geq h^2_{2,ij}$, together imply that $h_{2,i'j'}\geq h_{2,ij}$. 
\end{IEEEproof}
Now, let define $r^s_{ij} = \frac{1}{L}\sum_{k=1}^{L} r_{ij}[k]$. Using \eqref{f4}, we can rewrite $r^s_{ij}$ as
\begin{align}
\label{g2}
r^s_{ij}      = \frac{m \mu h_1 h_{2,ij}}{L} + n^s_{ij}
\end{align}
where $n^s_{ij}$ is a zero-mean Gaussian noise with variance 
\begin{align}
\label{g3}
\sigma^2_{s,ij} = \frac{m h_1 h_{2,ij} \sigma_s^2}{L^2} + \frac{\sigma^2_0}{L}.
\end{align}
As we can observe from \eqref{g3}, by increasing $L$, the variance of noise $n^s_{ij}$ approaches zero, and thus, for large values of $L$, one can reasonably neglect the effect of $n^s_{ij}$ and approximate $r^s_{ij}$ as $ r^s_{ij} \simeq \frac{m \mu h_1 h_{2,ij}}{L}$.
Given these assumptions and also according to the proposition, the vertical and horizontal intervals  on the APD array at which the center of the received optical beam is placed,  can be estimated as 
\begin{align}
\label{g1}
(\hat{i}',\hat{j}')= \underset{i,j\in\{1,...,N_a\}}{\operatorname{\text{arg~max~}}}    r^s_{ij}.
\end{align}
By feeding this information back to the CubeSat’s mechanical subsystem, such as a fast steering mirror, through a control message, the receiver can correct any orientation errors and realign itself with the direction of the incoming beam.
 
We note that \eqref{g1} gives an estimate of  the interval that the RVs $\theta_x$ and $\theta_y$ are located in, i.e., $\Big\{\frac{(i'-1-\frac{N_a}{2})w_a}{f_c}+\frac{w_f}{2f_c} <\theta_x < \frac{(i'-\frac{N_a}{2})w_a}{f_c}-\frac{w_f}{2f_c}\Big\}$, and 
$\Big\{\frac{(j'-1-\frac{N_a}{2})w_a}{f_c}+\frac{w_f}{2f_c} <\theta_y < \frac{(j'-\frac{N_a}{2})w_a}{f_c}-\frac{w_f}{2f_c}\Big\}$. Nevertheless, the accurate estimation of $\theta_x$ and $\theta_y$ requires the perfect knowledge of  the channel coefficients $h_1$ and $h_{2,ij}$, which is
beyond the scope of this work and can be an interesting future
direction.

\section{ Numerical Results And Discussion}
\label{numerical}

In this section, numerical results are presented in terms of BER to assess the performance of the proposed methods for both data detection and fine beam tracking. Additionally, Monte-Carlo simulations are conducted to validate the accuracy of the derived analytical expressions across various parameter values related to CubeSat's angular instabilities $\sigma_x^2$ and $\sigma_y^2$, transmit power $P_t$, the number of APDs in the array $N = N_a \times N_a$, and the observation window length $L$. Simulations are performed based on the practical values
of the parameters outlined in Table I \cite{ghassemlooy2012optical}.

\begin{table}
	\label{sys-parameters}
	\caption{System Parameters Used Throughout Simulations} 
	\centering 
	\begin{tabular}{c c c} 
		\hline\hline \\[-.5ex]
		Name & Parameter & Value \\ [.5ex] 
		\hline\hline \\[-1.2ex]
		Altitude difference                        & $ H_s - H_g $           &$ 400 $ km  \\[1ex] 
		Zenith Angle                        & $ \xi $           &$ 30^{\circ} $  \\[1ex] 
		APD Gain                        & $ G $           &$ 100 $  \\[1ex]
		Quantum Efficiency              & $ \eta $         & $ 0.7 $ \\[1ex]
		Wavelength                      &$ \lambda $      &$ 1550$ nm   \\[1ex]               
		Receiver Load                   & $ R_{l} $       & $ 1~k\Omega $ \\[1ex]                
		Receiver Temperature            &$ T_{r} $        & $ 300\degree~K$  \\[1ex] 
		Optical Filter Bandwidth            &$ B_{o} $        & $1$ nm  \\[1ex] 
		Spectral Radiance            &$ N_b(\lambda) $        & $10^{-4}$ Watts/${\rm cm}^2$-$\micro$m-srad  \\[1ex]              
		Ground Turbulence Level                     &$ C_n^2(0) $        & $ 10^{-13} $ $\textrm{m}^{-2/3}$ \\[1ex] 
		Bit Time                        &$ T_{b} $        & $ 10^{-9} $ \\[1ex]  
				Focal length               &$ f_c $     & $ 3 $ cm \\[1ex]               
		Aperture Radius                 & $ r $           & $ 3$ cm \\[1ex] 
		Beam Width           & $ w_z $       & $ 30$ m\\[1ex] 
		Background Power                & $ P_{b} $       & $ 1$ nW  \\[1ex]                
		Detector width                  &$ w_a $ & 250 $\micro$m \\[1ex]
		Dead-space width                  &$ w_f $ & 5 $\micro$m \\[.1ex]
		\hline\hline 
	\end{tabular}
\end{table}

\begin{figure}
	\centering
	\subfloat[] {\includegraphics[width=3.30 in]{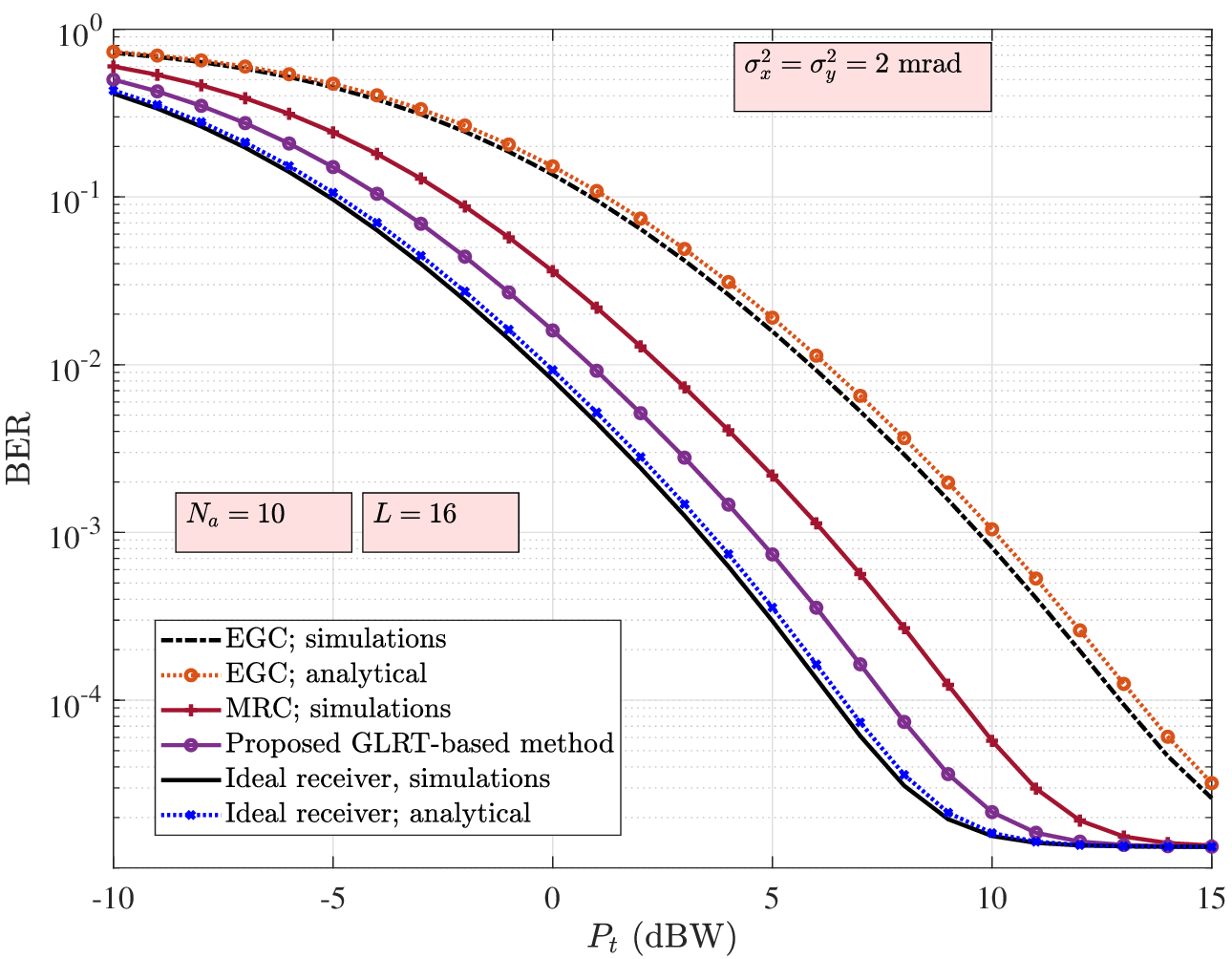}
		\label{pt1}
	}
	\hfill
	\subfloat[] {\includegraphics[width=3.30 in]{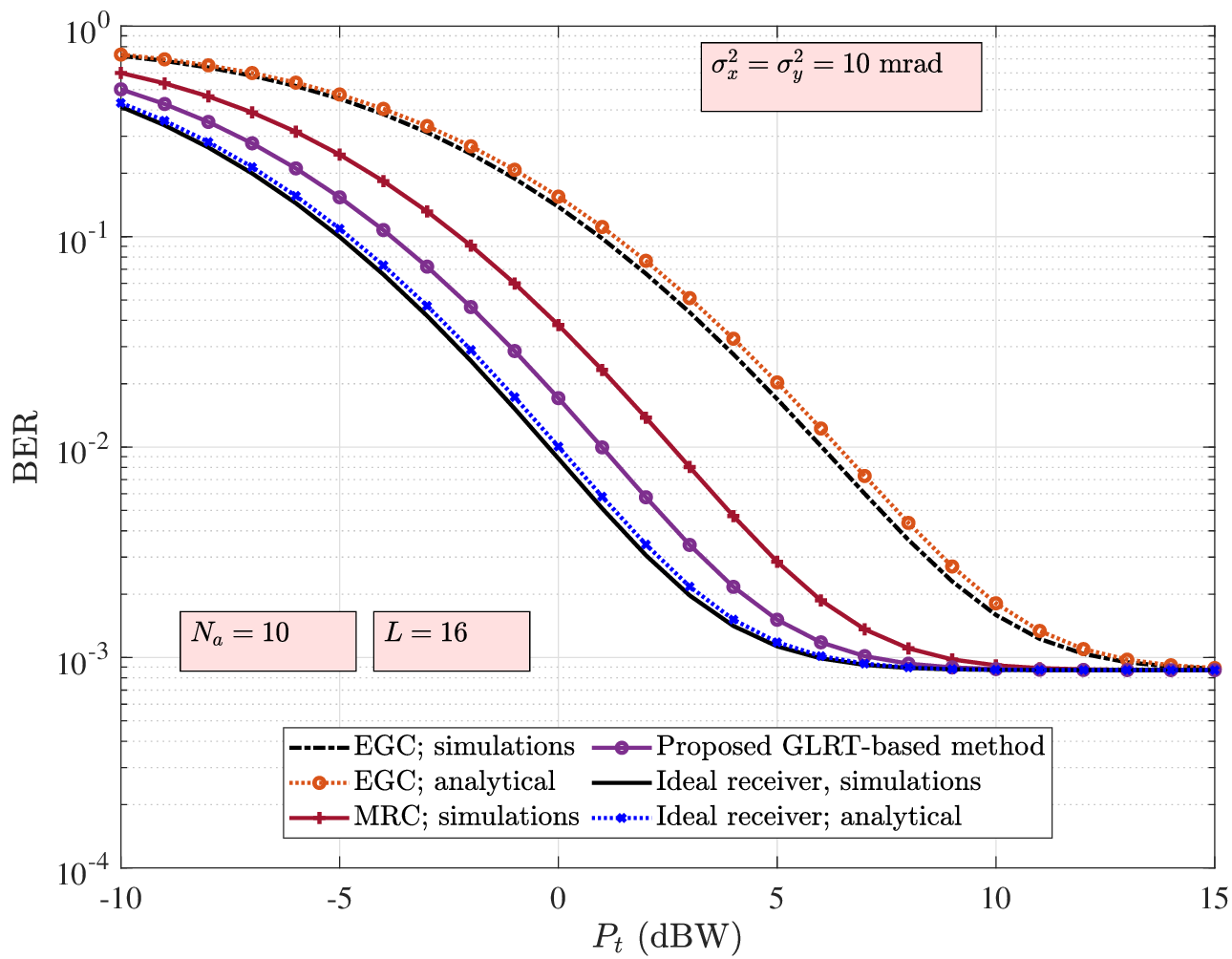}
		\label{pt2}
	}
	\caption{BER of the proposed GLRT-based detection method versus $P_t$ for two different angular instabilities. The case of employing  the EGC method and the ideal Rx are considered as benchmarks. }	\label{pt}
\end{figure}

We first investigate the performance of the proposed detection method in terms of BER. For evaluation, we also consider the BER results when the Rx perfectly knows the channel information (i.e., the ideal Rx), and when it employs the EGC and MRC techniques as benchmarks. Fig. \ref{pt} demonstrates BER versus $P_t$ for two different angular instabilities, i.e., when  $\sigma_x^2=\sigma_y^2= 2$ \textrm{mrad} (Fig. \ref{pt1}), and when  $\sigma_x^2=\sigma_y^2= 10$ \textrm{mrad} (Fig. \ref{pt2}). 
The results of Fig. \ref{pt} are obtained for $L=16$ and $N_a=6$. By comparing the results of Figs. \ref{pt1} and \ref{pt2}, one can readily observe that by increasing the CubeSat's angular instabilities, the link performance degrades significantly. Particularly, an error floor can be noticed in case of large angular instabilities and small Rx FoV due to the dominant effect of AoA fluctuations on the system performance.   Moreover, the results clearly prove the superiority of the proposed GLRT-based detection method compared with the EGC and MRC methods. Indeed, our proposed GLRT-based method outperforms the  aforementioned  combination techniques since it can exploit the continuity of the channel, and thus efficiently and accurately estimate the instantaneous channel coefficient to perform data detection. 
In addition, it is obvious that the MRC technique outperforms the EGC method at the expense of  high computational complexity.  Following the higher computational complexity and relatively poor performance, it can be concluded that the MRC technique is not a suitable method for data detection of the considered scenario.
Meanwhile, we notice a perfect match between the analytical and simulation-based results which validates the accuracy of our derived analytical expressions.

\begin{figure}[t]
	\begin{center}
		\includegraphics[width=3.3 in]{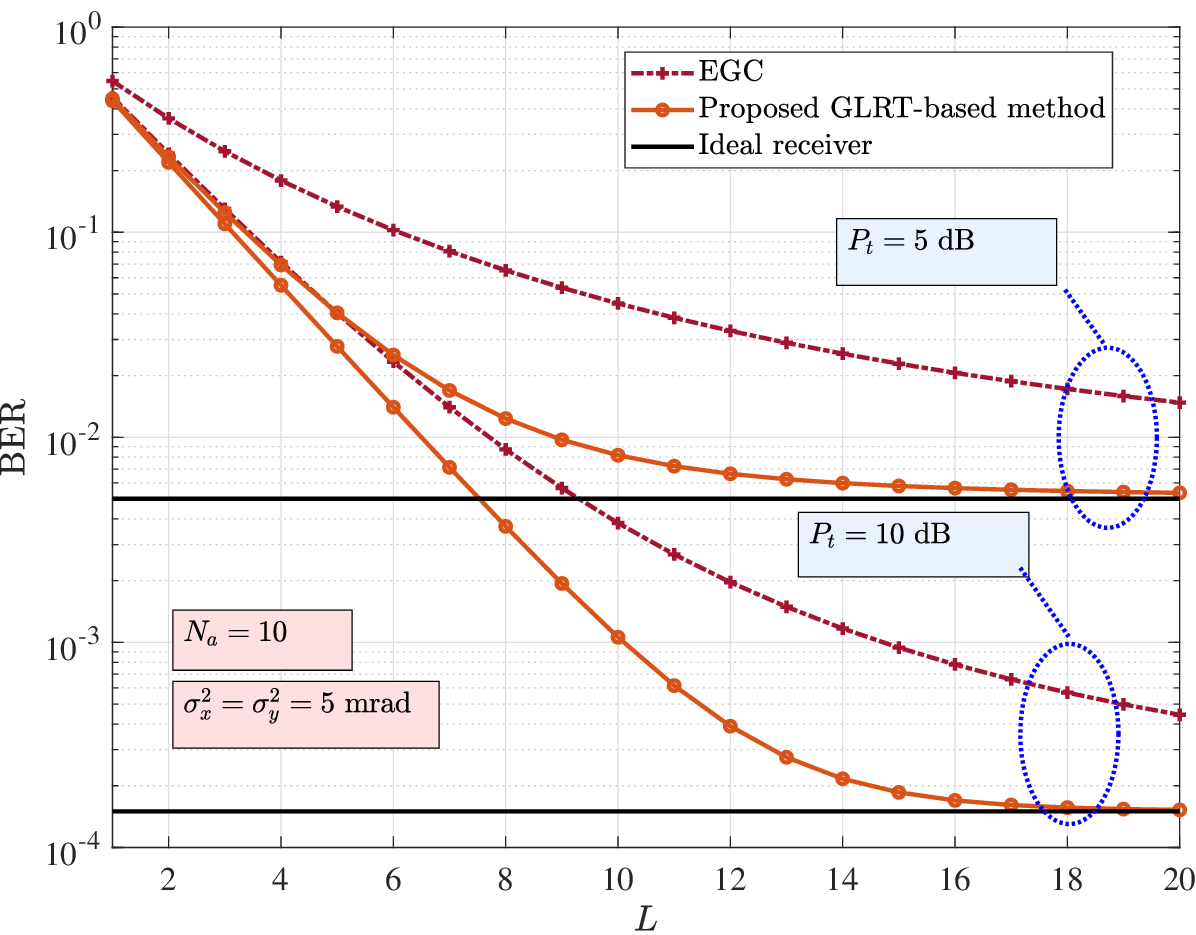}
		\caption{BER of the proposed GLRT-based detection method versus $L$ for two different values of $P_t$. The case of employing  the EGC method and the ideal Rx are considered as benchmarks.}
		\label{udddd1}
	\end{center}
\end{figure}

\begin{figure}[t]
	\begin{center}
		\includegraphics[width=3.3 in]{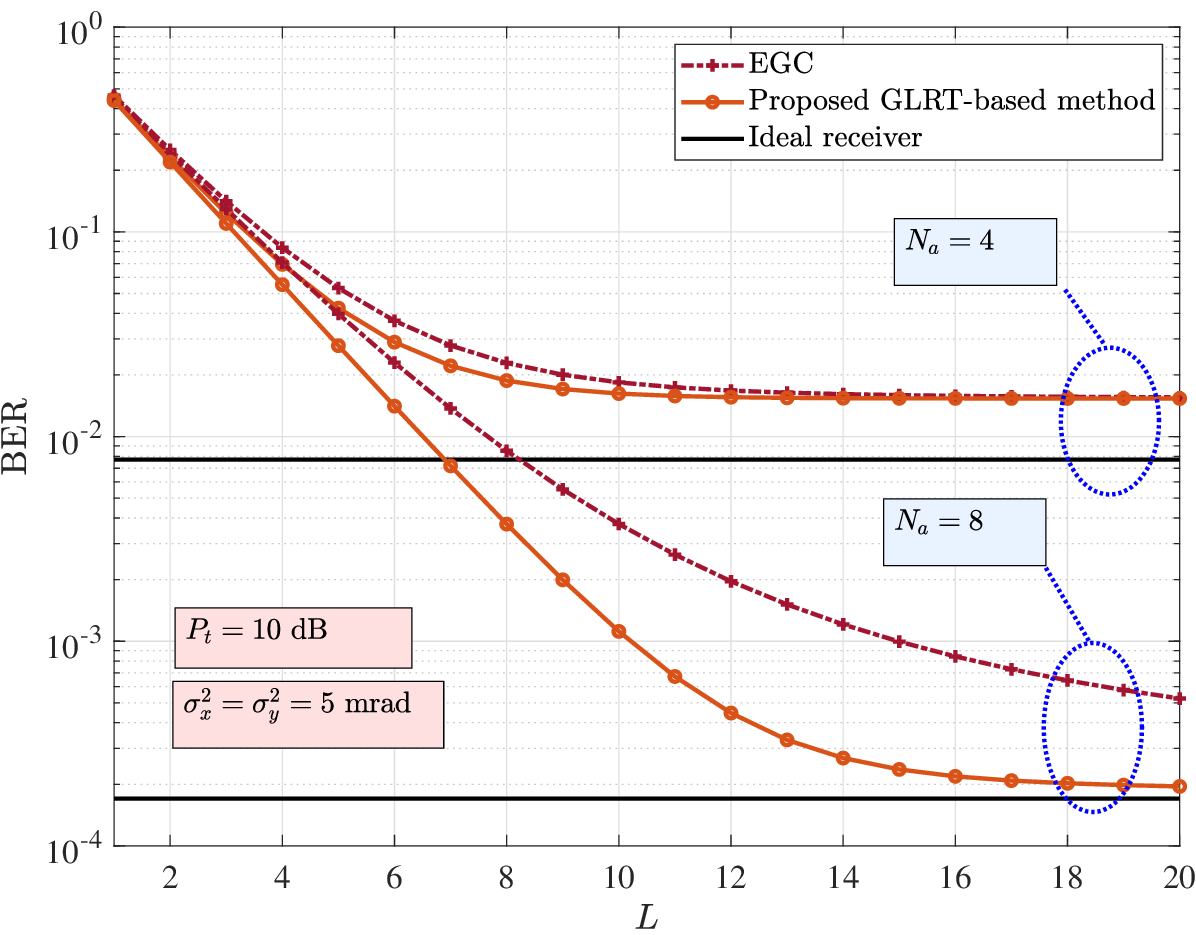}
		\caption{BER of the proposed GLRT-based detection method versus $L$ for two different values of $N_a$. The case of employing  the EGC method and the ideal Rx are considered as benchmarks.}
		\label{udddd1lef}
	\end{center}
\end{figure}
To have a deeper understanding about the effect of the length of observation window, $L$, on the link performance,
we have shown the BER
curves of the considered detection methods for $P_t$ of 5 dB and 10 dB versus $L$ in Fig. \ref{udddd1}. The results of this figure are obtained for $N_a=10$ and $\sigma_x^2=\sigma_y^2=5$ \textrm{mrad}. It is seen from Fig. \ref{udddd1} that the performance of the both proposed GLRT-based and EGC-based detection methods improve by increasing $L$. We note that this performance improvement comes at the cost of  increasing computational load as well as the detection delay. Hence, choosing an optimal
value for $L$ involves balancing a tradeoff between tolerable complexity/delay
and desirable BER. For instance, in the considered setup in Fig. \ref{udddd1} and when transmit power is equal to 5 dB, the lowest value of $L$ with which the system can achieve performance close to that of the ideal receiver is equal to $18$. Meanwhile, from the results of Fig. \ref{udddd1} we can observe that  the optimal value for $L$ changes by varying $P_t$. For instance, by increasing $P_t$ from 5 to 10 dB, the optimal value for $L$ changes from $18$ to $20$ to attain a lower BER. This can be justified by the fact that increasing $P_t$ leads to higher receiver SNR, subsequently reducing the BER of an ideal receiver. Therefore, for the proposed detection method, a larger value for $L$ is required to attain lower values of BER by mitigating detection and channel estimation errors. Besides $P_t$, another important parameter that can affect the optimal value of $L$ is the number of APDs in the detector array. To obtain more insight on this effect, we have presented the BER plots versus $L$ for two different values of $N_a$ in Fig. \ref{udddd1lef}. It can be seen form Fig. \ref{udddd1lef} that by increasing $N_a$ from $4$ to $8$, the optimal value of $L$ for the GLRT-based method increases from $13$ to $20$, and consequently, the BER decreases.

\begin{figure}[t]
	\begin{center}
		\includegraphics[width=3.3 in]{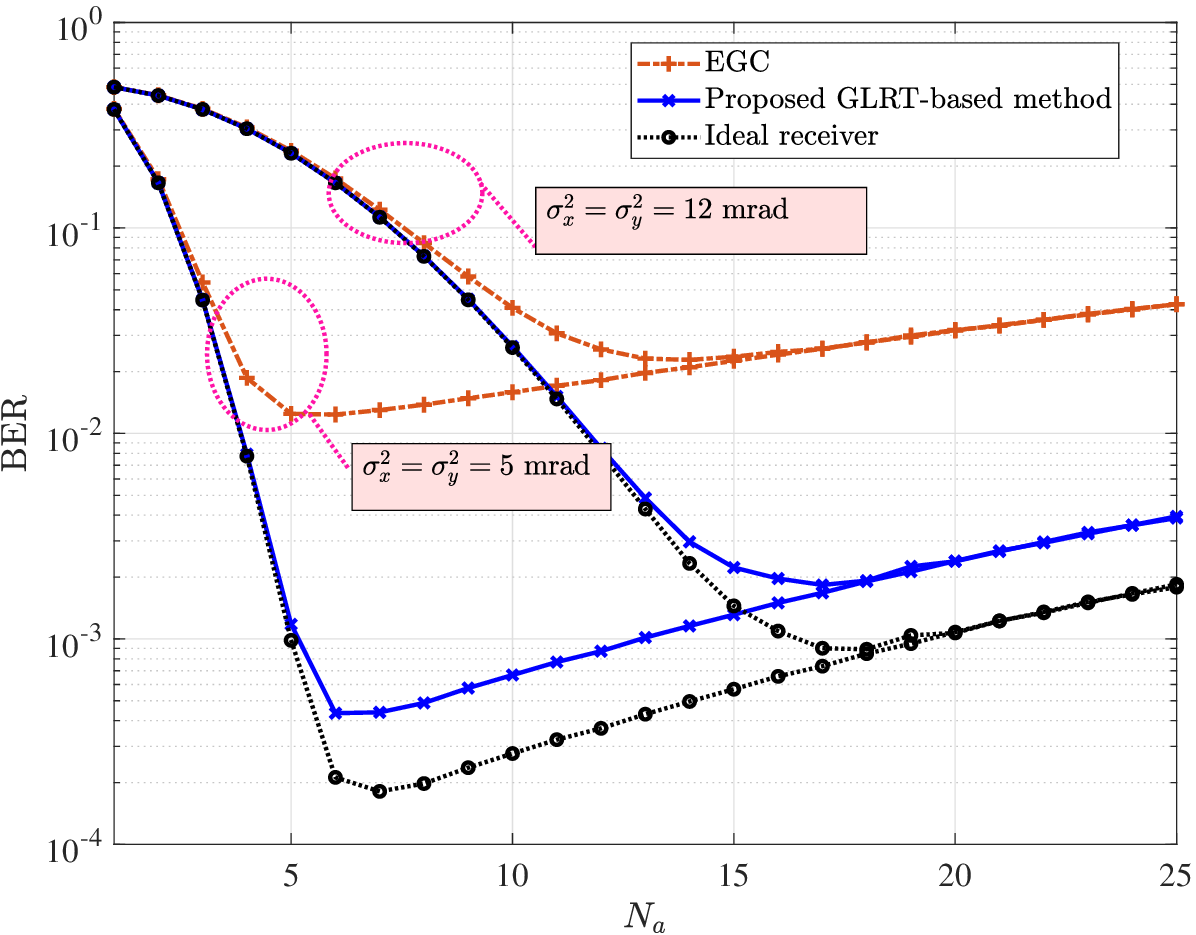}
		\caption{BER of the proposed GLRT-based detection method versus $N_a$ for two different values of angular instability. The case of employing  the EGC method and the ideal Rx are considered as benchmarks.}
		\label{udddd1lefrt}
	\end{center}
\end{figure}

The performance degradation due to the AoA fluctuations can be improved by increasing the Rx FoV via increasing the number of APDs (or equally $N_a$) in the detector array. Indeed, increasing $N_a$ makes a compromise between increased the Rx background noise level on one hand, and the decrease in the link interruption probability due to AoA fluctuations on the other hand. Moreover, since the computational complexity of the proposed detection method increases exponentially by increasing $N_a$, finding the optimal values for $N_a$ is essential to design such ground-to-CubeSat FSO links. Similar to $L$, the optimal value for $N_a$ is the minimum value that achieves performance close to the ideal receiver. To this end, we have plotted the BER curves of the different detection methods versus $N_a$ for two different values of angular instabilities in Fig. \ref{udddd1lefrt}. First, it can be seen from the figure that the performance of the system degrades when the degree of instabilities is varied form 5 to 12 $\textrm{~mrad}$, since, as expected,  higher degree of instabilities results in the 
larger value of AoA fluctuations (or equally, the incident optical beam is more likely to lie outside the Rx FoV). Second, by increasing the degree of instabilities form 5 to 12 $\textrm{~mrad}$, the optimal value for $N_a$ increases from $7$ to $18$ to compensate the effect of AoA fluctuations by employing a wider Rx FoV. Moreover, beyond an optimal point, further enlarging the detector size does not necessarily improve link reliability. Increasing the receiver FoV by enlarging the detector size entails capturing more desired transmit power along with undesired background noise. Consequently, beyond the optimal detector size (i.e., an optimal Rx field-of-view), background noise becomes dominant over the signal level, resulting in an increased BER.

Finally, as reducing computational complexity is often crucial to meet the SWaP requirements of CubeSats, we assess the significance of the proposed algorithm in terms of its processing time and computational load. It is demonstrated that the optimal value for $L$ is approximately 20 to achieve performance close to that of the ideal receiver at the target BER of $10^{-4}$. Notably, achieving a lower BER (i.e., lower than $10^{-4}$) will inevitably necessitate an increase in the optimal value for $L$. For the target BER of $10^{-4}$, the required time for searching through the entire search space in optimal detection is $2^{20}\times t_r = 73400320 t_r$, where $t_r$ represents the processing time unit for each realization and is directly related to the processing power of the processor unit in the receiver. However, our proposed method, which can achieve performance close to that of the ideal receiver, has a search complexity on the order of $L$, requiring only $20 t_r$ of processing time. This is significantly lower than the processing time of the exhaustive search method, resulting in reduced energy consumption and the ability to utilize simpler processing units in the payload of the CubeSat.

\section{Summary and Conclusion}
\label{conc}
In our study, we tackled the problem of data detection and fine beam tracking for ground-to-CubeSat FSO links using an array of APDs at the receiver. We developed a channel model tailored to this link and explored practical scenarios where both channel information and instantaneous beam position are unknown at the receiver. We proposed an efficient and practical data-aided channel estimation method based on the GLRT criterion and evaluated its performance under various conditions. Furthermore, we determined the center position of the beam spot on the APD array by comparing the output signals from the APDs. Our simulation results demonstrated that the proposed GLRT-based method achieves performance close to the ideal receiver while maintaining significantly lower complexity. This makes our method particularly suitable for CubeSats, which operate under stringent SWaP constraints. The reduced computational load and energy consumption allow for the use of simpler processing units, enhancing the feasibility of implementing advanced optical communication systems on CubeSats.

%


\end{document}